\documentclass[journal]{new-aiaa}
\usepackage[utf8]{inputenc}
\usepackage{textcomp}

\usepackage{float}

\usepackage{graphicx}
\usepackage{amsmath}
\usepackage{physics}
\usepackage[version=4]{mhchem}
\usepackage{siunitx}
\usepackage{longtable,tabularx}
\usepackage{multicol}
\usepackage{geometry}
\usepackage{algorithm}
\usepackage{subcaption}
\usepackage{algpseudocode}
\usepackage{environ}
\usepackage{todonotes}
\usepackage[normalem]{ulem} 

\NewEnviron{mcblock}{%
  \begingroup
  \color{blue}%
  \BODY
  \endgroup
}

\NewEnviron{mbblock}{%
  \begingroup
  \color{red}%
  \BODY
  \endgroup
}

\title{Reach-Avoid Differential game with Reachability Analysis for UAVs: A decomposition approach}

\author{
  Minh Bui 
  \footnote{PhD candidate, School of Computing Science, buiminhb@sfu.ca}
  \footnote{This work was during Minh's internship at Defense Reseach \& Development Canada (DRDC)}
} 
\affil{Simon Fraser University, Burnaby, BC V5A 1S6, Canada}

\author{
  Simon Monckton
  \footnote{Defense Scientist, DRDC Suffield Research Center, simon.monckton@forces.gc.ca}
}
\affil{Defense Research \& Development Canada (DRDC),  Medicine Hat, Alberta T1A 8K6, Canada}

\author{Mo Chen
\footnote{Associate Professor, School of Computing Science, mochen@cs.sfu.ca}
} 
\affil{Simon Fraser University, Burnaby, BC V5A 1S6, Canada}

\renewcommand\arraystretch{1.0}
\newtheorem{remark}{Remark}

\begin{document}

\maketitle


\begin{abstract}
    Reach-avoid (RA) games have significant applications in security and defense, particularly for unmanned aerial vehicles (UAVs).
    These problems are inherently challenging due to the need to consider obstacles, consider the adversarial nature of opponents, ensure optimality, and account for nonlinear dynamics.
    Hamilton-Jacobi (HJ) reachability analysis has emerged as a powerful tool for tackling these challenges; however, while it has been applied to games involving two spatial dimensions, directly extending this approach to three spatial dimensions is impossible due to high dimensionality.
    On the other hand, alternative approaches for solving RA games lack the generality to consider games with three spatial dimensions involving agents with non-trivial system dynamics.
    In this work, we propose a novel framework for dimensionality reduction by decomposing the problem into a horizontal RA sub-game and a vertical RA sub-game.
    We then solve each sub-game using HJ reachability analysis and consider second-order dynamics that account for the defender's acceleration.
    To reconstruct the solution to the original RA game from the sub-games, we introduce a HJ-based tracking control algorithm in each sub-game that not only guarantees capture of the attacker but also tracking of the attacker thereafter.
    We prove the conditions under which the capture guarantees are maintained.
    The effectiveness of our approach is demonstrated via numerical simulations, showing that the decomposition maintains optimality and guarantees in the original problem.
    Our methods are also validated in a Gazebo physics simulator, achieving successful capture of quadrotors in three spatial dimensions space for the first time to the best of our knowledge.

\end{abstract}

\section*{Nomenclature}

 {\renewcommand\arraystretch{1.0}
  \noindent\begin{longtable*}{@{}l @{\quad=\quad} l@{}}
      $\textbf{p}_{A}, \textbf{p}_{D}$  & X, Y, Z coordinates of attacker and defender respectively (meters)\\
      $\textbf{p}^h_{A}, \textbf{p}^h_{D}$ & X, Y coordinates of attacker and defender respectively (meters)\\
      $\mathbf{x}$& joint state vector of both attacker and defender \\
      $\mathbf{x}^{h}_{D}, \mathbf{x}^{h}_{A}$ & state vector in horizontal direction for defender and attacker \\
      $\mathbf{x}^{z}_{D}, \mathbf{x}^{z}_{A}$ & state vector in vertical direction for defender and attacker \\
      $V$ & Maximum distance value function between attacker and defender \\
      $\Phi$   & Signed-distance value function of reach-avoid game \\
      $T$ & Time horizon (seconds) \\
      $d_c$   & 3-dimensional game capture radius (meters) \\
      $d_h$  & horizontal capture radius (meters)\\
      $d_z$  & vertical capture radius (meters) \\
      $\mathcal{W}_{D, h}, \mathcal{W}_{D, z}$  & winning region for defender in horizontal, vertical game respectively\\
      $\mathcal{W}_{A, h}, \mathcal{W}_{A, z}$  & winning region for attacker in horizontal, vertical game respectively\\
      $k_x, k_y, k_z$ & proportional coefficients of directional velocities in $x, y, z$ direction respectively\\
      $\mathcal{T}$ & Target (goal) set\\
      $\mathbf{u}$ & Control input vector\\
      \multicolumn{2}{@{}l}{Subscripts}\\
      $D$ & Defender's attribute\\
      $A$ & Attacker's attribute\\
      $h$ & Horizontal's attribute\\
      $z$ & Vertical's attribute\\
  \end{longtable*}}

\section{Introduction}

Reach-avoid games are a class of differential games in which an agent (attacker) aims to reach a target set while avoiding predefined obstacles and an adversarial opponent (defender).
In this game, the attacker will try to arrive at the target goal region while the defender tries to intercept it before it can achieve its goal.
As technologies such as unmanned aerial vehicles (UAVs) and unmanned underwater vehicles (UUV) become more prevalent, reach-avoid games in three spatial dimensions have become more relevant than ever before.

Since the seminal work on differential games by Rufus Isaacs \cite{isaacs1965differential}, there has been substantial effort devoted to developing methods for solving these games.
Many approaches rely on geometric methods \cite{isaacs1965differential,Voronoi,oylerObs,GeometryJustified}, constructing dominance regions such as Voronoi cells and Apollonius circles for players with single integrator dynamics, where optimal trajectories are straight lines.
Other studies \cite{ruiAnlyticalBarrier,TwoDefAndOneAtt} construct these regions more analytically with convex regions for multi-agent target defense games.
In \cite{Garcia2,circularTargetDefense} the method of characteristics is used to solve a system of ordinary differential equations whose solutions satisfy the Hamilton-Jacobi-Isaac (HJI) partial differential equation.
While these methods have primarily focused on differential games with two spatial dimensions, they have also been extended to games with three spatial dimensions \cite{3DReachavoid1,3DReachavoid2,Rui3D2}, also for single integrator dynamics.
Three spatial dimensions introduce additional complexity by adding an extra spatial dimension for each attacker and defender, amplifying the challenge of analysis.

These existing methods often yield feasible results because of the homogeneous single-order integrator dynamics assumption for all players, which exhibit straight-line optimal trajectories.
However, this assumption on the model is not reflective of real-world autonomous systems, especially ones capable of high speed such as quadrotors, aircraft, etc. Consequently, the applicability of these approaches to realistic systems with high-dimensional state spaces remains unclear.

To overcome the constraints of simplified dynamics, reachability analysis has emerged as a powerful method, and involves directly solving the Hamilton-Jacobi (HJ) partial differential equation (PDE) to obtain viscosity solutions via level set methods \cite{Ianspaper}.
The biggest advantages of this approach are consideration of complex non-linear dynamical systems beyond single-integrator models, representation of complete winning regions, and guarantees of optimality in the resulting feedback strategies for each player \cite{HJforReachAvoid}.
HJ reachability analysis has also been applied effectively for safety verification of complex real-world robotics systems \cite{bansalHJVisual,annurev,ForcebasedLanding, leung2020infusingreachabilitybasedsafetyassurance,UnderwaterVehicle}, as well as multi-agent reach-avoid games for single integrator dynamics \cite{HighDimensionAdvances1,HighDimensionAdvances2, MaximumMatching}.
The main drawback of this approach is that the HJ PDE is solved numerically on a grid and hence has exponential time and memory complexity, rendering reachability analysis for systems with high-dimensional state spaces intractable.

In this paper, we set out to address this challenge and apply HJ reachability analysis for reach-avoid game in three spatial dimensions.
Our goal is to develop a tractable decomposition-based algorithm that approximates the solution to the original problem and
under the certain conditions, can guarantee a winning capture strategy for the defender.
In particular, we are interested in guaranteeing the performance of the defender in both capturing the attacker and avoiding the obstacles to win the game, while considering applicability of the numerical solution to fast real-world systems.
Such criteria require a higher-fidelity model of the defender that captures second-order dynamics in all three spatial dimensions.
Our paper employs HJ reachability to approximate the winning region for both the attacker and the defender.
For the first time, we obtain an approximative solution for the reach-avoid game in three spatial dimensions while accounting for second order dynamics of the defender, allowing us to validate the numerical solution on quadrotors simulated in Gazebo.
In addition to ensuring capture, the proposed method also enables tracking of the target 
thereafter, which paves the way towards real-world drone-on-drone interception.
This is achieved by the following contributions: 

\begin{enumerate} 
    \item Dynamics modeling simplification done in a conservative way: quadrotor dynamics involving a 12-dimensional (12D) state space are simplified into 6D double integrator dynamics for the defender and 3D single integrator dynamics for the attacker, giving the attacker advantage by over-approximating its physical capabilities. 
    \item Decomposition of horizontal and vertical spatial dimensions: The resulting problem involving a 9D state is broken down into a 6D horizontal sub-game and 3D vertical sub-game that can be \textbf{tractably solved} using HJ reachability.
    \item The reach-track control algorithm, allowing solutions of the subgames to be combined while maintaining guarantees on capture in the original game
    \item Theoretical analysis of guarantees provided by our methods
    \item Numerical and physics-based drone-on-drone interception simulations in Gazebo that validate our theoretical results
\end{enumerate}


\section{Preliminaries}
\subsection{Hamilton-Jacobi Reachability Analysis}

Let $t$ be time and $\mathbf{x} = \begin{bmatrix} \mathbf{x}_{D} \\ \mathbf{x}_{A} \end{bmatrix} \in \mathbb{R}^{N_A + N_D}$ be the joint state space of both the attacker and defender.
The evolution of this joint system over time is described by a system of ordinary differential equations (ODE) \cite{HJTutorial}:  \begin{equation} \label{eq:ode} \begin{aligned} \dot{\mathbf{x}} = \dfrac{d\mathbf{x}(t)}{dt} = f(\mathbf{x}(t), \mathbf{u}_{D}(t), \mathbf{u}_{A}(t)), \quad \text{with } t \in [0, T] \end{aligned} \end{equation} where $T$ is the duration of the reach-avoid game.
In addition, the control inputs $\mathbf{u}_{D}(\cdot), \mathbf{u}_{A}(\cdot)$ are for the defender and attacker respectively.
We assume that the system dynamics $f$ is uniformly continuous, bounded, and Lipschitz continuous in $\mathbf{x}$ for fixed $ \mathbf{u}_{D}(\cdot)$ and $\mathbf{u}_{A}(\cdot)$.
Hence, given $\mathbf{u_{D}(\cdot)}$ and $\mathbf{u_{A}}(\cdot)$, there exists a unique trajectory that satisfies Eq.
\eqref{eq:ode}, which is denoted as
$\zeta^{\mathbf{u_A}(\cdot),  \mathbf{u_D}(\cdot)}_{\mathbf{x}, t'}(t)$  for $t > t'$ with the initial condition $\zeta^{\mathbf{u_A}(\cdot),  \mathbf{u_D}(\cdot)}_{\mathbf{x}, t'}(t') = \mathbf{x}$.

\begin{figure}[H]
    \centering
    \includegraphics[scale=0.65]{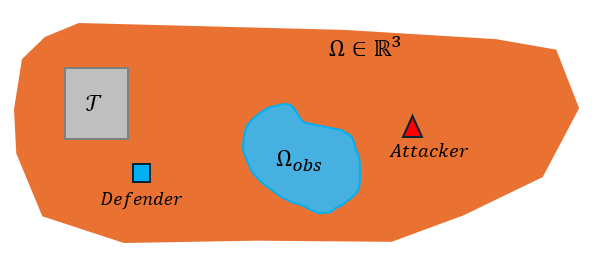}
    \caption{In three spatial dimensions, the defender needs to capture the attacker before it reaches $\mathcal{T}$ while avoiding $\Omega_{obs}$.}
    \label{fig:problem_assumption}
\end{figure}

The attacker \textbf{wins} when the joint state belongs to the following reach set, which represents the joint states where given their respective optimal policies, the attacker is able to arrive at the target region $\mathcal{T}$ or the defender inevitably hits the obstacle $\Omega_\text{obs}$:

\begin{equation}
    \label{eq: original_reach_set}
    \mathcal{R}=  \left\{\mathbf{x} \mid \mathbf{x}_{A} \in \mathcal{T}  \right\} \cup \left\{ \mathbf{x} \mid \mathbf{x}_D \in \Omega_\text{obs}\right\}
\end{equation}

On the other hand, the defender \textbf{wins} when $\mathbf{x}$ arrives at the following avoid set:
\begin{equation}
    \label{eq: original_avoid_set}
    \mathcal{A}= \left\{ \mathbf{x} \mid\left\|\boldsymbol{p}_{A}- \boldsymbol{p}_D\right\|_2 \leq d_c \wedge \mathbf{x}_{A} \notin \mathcal{T} \right\} \cup\left\{\mathbf{x} \mid \mathbf{x}_{A} \in \Omega_\text{obs}\right\},
\end{equation}

\noindent which represents the joint states where the attacker is captured outside the target set $\mathcal{T}$ or hits the obstacles.
Note that reach and avoid sets defined represent the desirable and undesirable states respectively, for the attacker.
Numerically, reach set $\mathcal{R}$ is represented by the sub-zero level of some function $l(\mathbf{x})$
while avoid set $\mathcal{A}$ is
represented by the super-zero level of some function $g(\mathbf{x})$:
\begin{equation} \begin{aligned}
     \label{eq: level_set_representation}
     \mathcal{R} =\left\{\mathbf{x} \in \mathbb{R}^{N_A + N_D} \mid l(\mathbf{x}) \leq 0\right\} \\ 
     \mathcal{A} =\left\{\mathbf{x} \in \mathbb{R}^{N_A + N_D} \mid g(\mathbf{x}) > 0\right\} 
\end{aligned} \end{equation}

To construct these function $l(x)$ and $g(x)$, we first need a function representation for each of the term
in Eq. \eqref{eq: original_reach_set} and \eqref{eq: original_avoid_set} and 
then combine them together.
To do so, we first define the signed distance function representation of a set $C$:
\begin{equation}
    S(\mathbf{x}, C) =
    \begin{cases}
        -d(\mathbf{x}, \partial C), & \text{if } \mathbf{x} \in C \\
        d(\mathbf{x}, \partial C), & \text{if } \mathbf{x} \notin C
    \end{cases}
\end{equation}
where $d(\cdot)$ is the nearest distance between point $x$ and the set boundary $\partial C$.
The points outside the set are positive, while those inside are negative.
Now we can construct $l(\mathbf{x})$ as follows:
\begin{equation}
    \begin{aligned}
    l_{\mathcal{T}}(\mathbf{x}) = S(\mathbf{x}_A, \mathcal{T}), \quad l_{\Omega_\text{obs}}(\mathbf{x}) = S(\mathbf{x}_D, \Omega_\text{obs}) \\
    l(\mathbf{x}) = \min\{l_{\mathcal{T}}(\mathbf{x}), l_{\Omega_\text{obs}}(\mathbf{x})\}
    \end{aligned}
\end{equation}
If $l(\mathbf{x}) \leq 0$, the joint state $\mathbf{x}$ is inside the reach set $\mathcal{R}$ and outside otherwise.
In general, the union operation over sets is represented by taking the minimum of the
sign-distance function of these sets, while intersection of sets is represented by taking the maximum.
By sequentially applying these operations, one can represent arbitrary complex set shapes, a technique
widely used in computer graphics \cite{IntroImplicitSurfaces}.
Similar to $l(x)$, we can construct $g(x)$ to represent the avoid set $\mathcal{A}$ in Eq. \eqref{eq: original_avoid_set} as follows:
\begin{equation}
    \begin{aligned}
    g_{\text{capture}}(\mathbf{x}) = \left\|\boldsymbol{p}_{A}- \boldsymbol{p}_D\right\|_2 - d_c, \quad g_{\notin \mathcal{T}}(\mathbf{x}) = -S(\mathbf{x}_A, \mathcal{T}), \quad g_{\Omega_\text{obs}}(\mathbf{x}) = S(\mathbf{x}_A, \Omega_\text{obs}) \\
    g(\mathbf{x}) = -\min\{\max\{g_{\text{capture}}(\mathbf{x}), g_{\notin \mathcal{T}}(\mathbf{x}) \},  g_{\Omega_\text{obs}}(\mathbf{x})\}
    \end{aligned}
\end{equation}
Note in the above that $g(\mathbf{x})$ is negated at the end to ensure that $g(\mathbf{x}) > 0$ when $\mathbf{x} \in \mathcal{A}$.


Next, let's consider the functional outcome of the game \cite{Evans1983DifferentialGA}:

\begin{equation} 
    \label{eq:V_z_t}
     \mathcal{V}(\mathbf{x}, t, \mathbf{u_{D}}(\cdot), \mathbf{u_{A}}(\cdot)) = \min_{\tau \in[t,T]} \max \left\{l( \zeta^{\mathbf{u_A}(\cdot),  \mathbf{u_D}(\cdot)}_{\mathbf{x}, t}(\tau)), \max_{s\in [t, \tau]} g(\zeta^{\mathbf{u_A}(\cdot),  \mathbf{u_D}(\cdot)}_{\mathbf{x}, t}(s)) \right\} 
\end{equation}

In the above, the innermost max considers, at each time $\tau$, 
the greatest value of $g$ so far from time $t$ to $\tau$; if any value of $g$ is positive,
it means the joint state has entered the avoid set $\mathcal{A}$ at some time $s \in [t, \tau]$.
Furthermore, the inner expression $\max \left\{l( \zeta^{\mathbf{u_A}(\cdot),  \mathbf{u_D}(\cdot)}_{\mathbf{x}, t}(\tau)), 
\max_{s\in [t, \tau]} g(\zeta^{\mathbf{u_A}(\cdot),  \mathbf{u_D}(\cdot)}_{\mathbf{x}, t}(s)) \right\} $
 is less than or equal to zero if and only if the joint state 
enters the reach set $\mathcal{R}$ at time $\tau$ and never enter the avoid set $\mathcal{A}$ at all time $s \in [t, \tau]$
(using implicit function definitions of sets in Eq. \eqref{eq: level_set_representation}).
If this is true at any point for $\tau \in [t, T]$, then the attacker wins the game, which is reflected in 
the outermost min operation in Eq. \eqref{eq:V_z_t}. In this case, the sub-zero level set of $\mathcal{V}$
indicates the winning region for attacker over time horizon $T$;
for that reason, the attacker will minimize $\mathcal{V}$ 
while the defender will maximize $\mathcal{V}$.

Assuming that Isaac's condition is satisfied for the given dynamics, the optimizing order of each agent does not matter and the value of the game is well-defined as follows:

\begin{equation}
    \label{eq:game_value}
    \Phi(\mathbf{x}, t) := \inf_{\mathbf{u_{A}(\cdot)}} \sup_{\mathbf{u_{D}}(\cdot)} \mathcal{V}(\mathbf{x}, t, \mathbf{u_{D}}(\cdot), \mathbf{u_{A}}(\cdot))
\end{equation}
It has been shown in \cite{TimeVaryingRA} that the value function $\Phi(\textbf{x}, t)$ is the unique viscosity solution to the following HJI variational inequality with a final time condition:

\begin{align} \label{eq:HJI}
    \max\left\{\min \left\{\frac{\partial \Phi}{\partial t} + H \left(\mathbf{x}, \frac{\partial \Phi}{\partial \mathbf{x}}\right), l(\mathbf{x})-\Phi(\mathbf{x}, t) \right\},  g(\mathbf{x})-\Phi(\mathbf{x}, t)\right\} & =0, \quad t \in[0, T] \\ 
    \Phi(\mathbf{x},  T ) &= \max \left\{l(\mathbf{x}), g(\mathbf{x}) \right\}
\end{align}

\noindent where the optimal Hamiltonian $H$ is calculated as

\begin{equation} \label{eq:Hamiltonian}
    H \left(\mathbf{x}, \frac{\partial \Phi}{\partial \mathbf{x}}\right)= \max_{ \mathbf{u_D}} \min_{\mathbf{u_A}}  \left(\frac{\partial \Phi}{\partial \mathbf{x}}\right)^{\top} f(\mathbf{x}, \mathbf{u_A}, \mathbf{u_D})
\end{equation}

\noindent where the min-max order does not matter if the Isaac's condition is satisfied.
The above variational inequality can be efficiently solved using open-source numerical software libraries \cite{optimized, toolboxLS}.

As we let $T \rightarrow \infty$ the value function converges and we will denote it as $\Phi_{\infty}(\mathbf{x})$.
Once the value function $\Phi$ is obtained, the state feedback optimal controls for each player can be computed using the spatial derivatives information of the value of the game:

\begin{subequations} \label{eq:state_feedbackcontrol}
    \begin{align}
         & \mathbf{u}^*_{D}(\mathbf{x})=\arg  \max _{\mathbf{u_D} } \min _{\mathbf{u}_A}  \left(\frac{\partial \Phi}{\partial \mathbf{x}}\right)^{\top} f(\mathbf{x}, \mathbf{u}_{A}, \mathbf{u}_{D}) \\
         & \mathbf{u}^*_{A}(\mathbf{x})=\arg \min _{\mathbf{u}_{A} } \left(\frac{\partial \Phi}{\partial \mathbf{x}}\right)^{\top} f\left(\mathbf{x}, \mathbf{u}_{A}, \mathbf{u}^*_{D}\right)
    \end{align}
\end{subequations}

Using the computed $\Phi_{\infty}$, we can also obtain the winning region for both the attacker and defender.
The winning regions for the defender and attacker are defined below and can be numerically obtained using the super-zero and sub-zero level set of $\Phi_{\infty}(\mathbf{x} )$, respectively:

\begin{subequations}
    \begin{align}
        \mathcal{W}_A & :=\{\mathbf{x} \mid \exists \tau \in[0, +\infty), \exists \mathbf{u}_{A}(\cdot) \in \mathbb{U_A}, \forall \mathbf{u}_{D}(\cdot) \in \mathbb{U_D}, \zeta^{\mathbf{u_A}(\cdot),\mathbf{u_D}(\cdot)}_{\mathbf{x}, 0}  (\tau) \in \mathcal{R}, \\
                      & \qquad\forall \kappa \in[0, \tau], \zeta^{\mathbf{u_A}(\cdot),  \mathbf{u_D}(\cdot)}_{\mathbf{x}, 0}(\kappa) \notin \mathcal{A} \} \nonumber                                                                                               \\
                      & =  \{\mathbf{x} \mid \Phi_{\infty}(\mathbf{x}) \leq 0\}                                                                                                                                                                                    \\
        \mathcal{W}_D & :=\{\mathbf{x} \mid \exists \tau \in[0, +\infty), \exists \mathbf{u}_{D}(\cdot) \in \mathbb{U_D}, \forall \mathbf{u}_{A}(\cdot) \in \mathbb{U_A}, \zeta^{\mathbf{u_A}(\cdot),\mathbf{u_D}(\cdot)}_{\mathbf{x}, 0}  (\tau) \in \mathcal{A}, \\
                      & \qquad\forall \kappa \in[0, \tau], \zeta^{\mathbf{u_A}(\cdot),  \mathbf{u_D}(\cdot)}_{\mathbf{x}, 0}(\kappa) \notin \mathcal{R} \} \nonumber                                                                                               \\
                      & =  \{\mathbf{x} \mid \Phi_{\infty}(\mathbf{x}) > 0\}
    \end{align}
\end{subequations}

Although directly solving Eq.~\eqref{eq:HJI} provides a very general approach for computing the winning regions for different system dynamics, it relies on numerical computation on discretized grids that scale exponentially with the number of states and agents.
Currently, the limit of solving the variational inequality \eqref{eq:HJI} is only possible for joint systems $\mathbf{x}$ of at most 6-7 dimensions.

\subsection{Quadrotor dynamics}

Even though our proposed methods are not targeted towards any particular physical system, our paper will adopt quadrotors as the main application of our algorithmic framework.
We will adhere to the standard system dynamics for quadrotors.
This is the basic dynamics of a quadrotor that models the effect of the four throttles on the positional and rotational states of the quadrotor, which is used for quadrotor simulation in Gazebo for our experiment section.
\begin{equation} \label{eq:quadrotor_dynamics}
    \begin{bmatrix}
        \dot{x} \\ \dot v_x \\ \dot{y} \\ \dot v_y \\ \dot{z} \\ \dot v_z \\ \dot\omega_x^b \\ \dot\omega_y^b \\\dot\omega_z^b \\ \dot{\phi} \\ \dot{\theta} \\ \dot{\psi}
    \end{bmatrix}
    =
    \begin{bmatrix}
        v_x \\ 
        \frac{1}{m} \sum^{4}_{i=1} T_{i} \left(\cos\phi\sin\theta\cos\psi + \sin\phi\sin\psi \right) \\
        v_y \\  
        \frac{1}{m} \sum^{4}_{i=1} T_{i} \left( \cos\phi\sin\theta\sin\psi - \sin\phi\cos\psi \right)  \\ 
        v_z \\ 
        \frac{1}{m} \sum^{4}_{i=1} T_{i}\left(\cos\phi \cos\theta \right) - g \\ 
        \frac{l}{2I_x}(T_1 -T_2 -T_3+T_4) + \frac{1}{I_x}\omega_y^b\omega_z^b(I_y - I_z) \\ 
        \frac{l}{2I_y}(-T_1 -T_2 +T_3+T_4) + \frac{1}{I_y}\omega_x^b\omega_z^b(I_z - I_x) \\ 
        \frac{l}{2I_z} \tau_{\text{drag}}^b+ \frac{1}{I_z}\omega_x^b\omega_y^b(I_x - I_y) \\ 
        \omega_x^b + \omega_y^b\sin\phi\tan\theta + \omega_z^b \cos\phi\tan\theta \\ 
        \omega_y^b \cos\phi -\omega_z^b\sin\phi\\ 
        -\omega_y^b \dfrac{\sin\phi}{\cos\theta} +  \omega_z^b\dfrac{\cos\phi}{\cos\theta}
    \end{bmatrix}
\end{equation}

\noindent where $T_i$ is the thrust force of each rotor, $\omega_x^b, \omega_y^b, \omega_z^b$ is the rotational velocity with respect to the body frame, and $\phi, \theta, \psi$ are the roll, pitch, yaw in the global coordinate frame.
The above equation describes the rates of changes of states for only one drone.
In the reach-avoid game, typically we need to concatenate the states of both defender and attacker quadrotors together, which would result in a 24D system.
Because of such high-dimensionality, it is impossible to directly apply reachability analysis to this system, and we cannot use the full dynamics for analyzing the reach-avoid game.
In the next section, we will go into more detail about the high-level modeling that approximates the behaviors of both attacker and defender drones that can reduce the total number of states to enable tractable analysis.

\section{Dimensionality Reduction}

\subsection{High-level Modeling of Quadrotors}
In differential reach-avoid games, since we are mainly interested in the captured states that depend solely on the spatial coordinates and not the rotational angles.
This allows our high-level system modeling not to consider the full 12D system but rather the dynamics of the subset of ``slow" system components, which can be good approximations while allowing tractable game theoretical analysis and optimal control computation.
For simplicity, we will only consider the movement mode in which quadrotors exhibit translational degree of freedom in spatial space.
Such mode can be achieved by sending directional velocity commands in all 3 directions with respect to a local coordinate frame to low-level controllers \cite{px4_autopilot, ardupilot}.
Not just quadrotors, but any other UAV systems that exhibit translation freedom in the $x$-, $y$-, and $z$-axes can also be analyzed using high-level system modeling discussion in this section.

We are interested in modeling a fast defender whose maximum speed can be twice as fast as a slower attacker in both horizontal and vertical directions.
In this paper, we will model the fast defender using a double integrator dynamics while a single integrator is used for the slower attacker:

\begin{equation} \label{eq:3D_modeling}
    \begin{aligned}
        \dot{\mathbf{x}}_{D} =
        \begin{bmatrix}
            \dot{\mathbf{p}}_{D} \\ \dot{\mathbf{v}}_{D}
        \end{bmatrix}
        =
        \begin{bmatrix} \dot{x}_{D} \\ \dot{y}_{D} \\ \dot{z}_D \\ \dot{v}^{x}_D \\ \dot{v}^{y}_D \\  \dot{v}^z_D
        \end{bmatrix}
        =
        \begin{bmatrix}
            v^{x}_D \\ v^{y}_D \\ v^{z}_D \\ k_x(v^{x, C}_{D} - v^{x}_D)\\ k_y(v^{y, C}_{D} - v^{y}_D) \\ k_z(v^{z, C}_{D} - v^{z}_D)
        \end{bmatrix},
        \qquad
        \dot{\mathbf{x}}_{\mathbf{A}} = \dot{\mathbf{p}}_{\mathbf{A}} =
        \begin{bmatrix}
            \dot{x}_{A} \\ \dot{y}_{A} \\ \dot{z}_A
        \end{bmatrix}
        =
        \begin{bmatrix}
            v^{x, C}_{A} \\ v^{y, C}_{A} \\v^{z, C}_{A}
        \end{bmatrix}
    \end{aligned}
\end{equation}

\noindent where $k_x, k_y, k_z$ are the proportional coefficients for each direction, and $\mathbf{u}_{D} = (v^{x, C}_{D}, v^{y, C}_{D}, v^{z, C}_{D}), \mathbf{u_A} = (v^{x, C}_{A}, v^{y, C}_{A}, v^{z, C}_{A})$ are the control inputs for the defender and attacker respectively, each of which can be treated as the velocity commands sent to lower level controllers such as PX4 or Ardupilot \cite{ardupilot, px4_autopilot}.
Since we are interested in controlling a faster quadrotor defender, using a double integrator dynamics can more accurately describe the system capability, which is also very important for tracking the attacker and obstacle avoidance.
Realistically, velocity cannot be changed instantaneously at higher speeds, especially with optimal bang-bang controllers.
This is reflected in the defender's dynamics, where velocity command inputs cannot instantaneously turn into velocity state but instead affect the acceleration; effectively, acceleration approaches zero as the velocity gets close to the velocity command.
A more agile and maneuverable attacker is modeled using the single integrator dynamics, where the velocity is instantaneously set as the velocity command.
This dynamics overestimates the attacker's actual physical capability at higher speed, resulting in a more conservative yet robust control solution for the defender when the reach-avoid game is solved.

The constraints imposed on the control inputs for each player are as follows:

\begin{subequations} \label{eq:constraints}
    \begin{align}
        \sqrt{(v^{x,C}_D)^2 + (v^{y, C}_D)^2}  \leq \mathbf{U}^{h}_{D}  \\
        \sqrt{(v^{x, C}_A)^2 + (v^{y, C}_A)^2}  \leq \mathbf{U}^{h}_{A} \\
        \abs{v^{z, C}_D } \leq   \mathbf{U}^{z}_{D}                     \\
        \abs{v^{z, C}_A } \leq   \mathbf{U}^{z}_{A}
    \end{align}
\end{subequations}

\noindent where $\mathbf{U}^{h}_{D}, \mathbf{U}^{h}_{A}$ are the maximum horizontal speed, and $\mathbf{U}^{z}_{D}, \mathbf{U}^{z}_{A}$ are the maximum vertical speed for the defender and attacker respectively.
Since we give more agility advantage to the attacker by modeling it as a single integrator, a higher 
maximum speed capability for the defender compared to that of attacker can be chosen to balance the modeling choices,
ensuring non-empty winning region for the defender when the reach-avoid game is solved. In general, any speed ratio
can be used for our method presented in this paper.
Note that the limits of horizontal and vertical velocity commands are specified separately, which aligns well with the control architecture of the low-level controller provided by such as Ardupilot or PX4 \cite{ardupilot, px4_autopilot}.
Such decoupling in constraints between the vertical and horizontal velocity components will be convenient for our theoretical approach, which is different from existing work that imposes constraints on three inputs in all directions together.

Compared to the 24D joint quadrotor dynamics in Eq.
\eqref {eq:quadrotor_dynamics} for both attacker and defender, the total joint dynamics in Eq. \eqref {eq:3D_modeling} for both players is 9-dimensional.
This joint 9D dynamics, however, still remain out-of-reach for numerical computation using reachability toolbox \cite{optimized}.
In the next few sub-sections, we will break down this 9D system further into smaller sub-systems used in sub-games for horizontal and vertical direction, allowing tractable analysis.

\subsection{Capture Set Decomposition}
Typically in reach-avoid or pursuit-evasion games, the attacker is considered captured by the defender when the $L_2$ norm of the their relative distances is less than some threshold $d_c$.
Specifically in existing works \cite{3DReachavoid1, 3DReachavoid2, Rui3D2}, the capture condition is $ || \mathbf{p_D} - \mathbf{p_A} ||_{2} \leq d_c$ where $\mathbf{p_D} = \begin{bmatrix} x_D \\ y_D \\ z_D \end{bmatrix}$,  $\mathbf{p_A} = \begin{bmatrix} x_A \\ y_A \\ z_A \end{bmatrix}$ are the 3D positions of the defender and attacker with respect to a local coordinate frame $O$, respectively.

Instead of using this condition, we break this down into two sub-conditions that need to be satisfied \textbf{simultaneously} for a defender to \textbf{capture} the attacker :

\begin{subequations} \label{eq:capture_condition}
    \begin{align}
        \sqrt{(x_A - x_D)^2 + (y_A - y_D)^2 } \leq d_{h} \label{eq:horizontal_condition} \\
        \sqrt{(z_A - z_D)^2} \leq d_{z} \label{eq:vertical_condition}
    \end{align}
\end{subequations} 

Given attacker's position $\mathbf{p_A}$, the geometrical captured space is a cylinder of height $d_z$ and base radius $d_{h}$.
In the next section, we will show that the above captured condition will facilitate decomposition of the reach-avoid game into sub-games, where each game is independently defined by the corresponding capture condition above, respectively.

\subsection{Reach-Avoid Game Decomposition}

In this section, we will decompose the reach-avoid game in three spatial dimension into two sub-games: a horizontal game and a vertical game.
In the horizontal sub-game, we will consider a sub-system of Eq.~\eqref{eq:3D_modeling} that contains only dynamical components that are in x-y coordinates.
Similarly, for the vertical sub-game, only vertical components are considered. Before then, we will make the following assumption:

\textit{\textbf{Assumption:} the target region $\mathcal{T}$ and obstacles $\Omega_\text{obs}$ depend only on the horizontal coordinates and not on the vertical coordinates.}

This assumption means that the projection of the obstacle set and the goal set from three spatial dimensions into the horizontal spatial dimensions and vice versa is perfect, with no information loss in the process.
The target region depending only on horizontal coordinate is also a practical assumption for no-fly zone with low-altitude drone operation.
In air traffic control, this assumption can be considered for controlled airspace zones within its
limited altitude range. 
Within the altitude range of an airspace zone \cite{FAA_airspace_zone}, modeling these zones as target set or obstacles depending only on
horizontal coordinates still provide a meaningful approximation of the structure of the reach-avoid game.
Using the above assumption, together with capture set decomposition from previous section, we will now formalize the two sub-games and compute each player's optimal strategies,
each of which is computationally tractable.

\begin{figure}[H]
    \centering
    \includegraphics[scale=0.58]{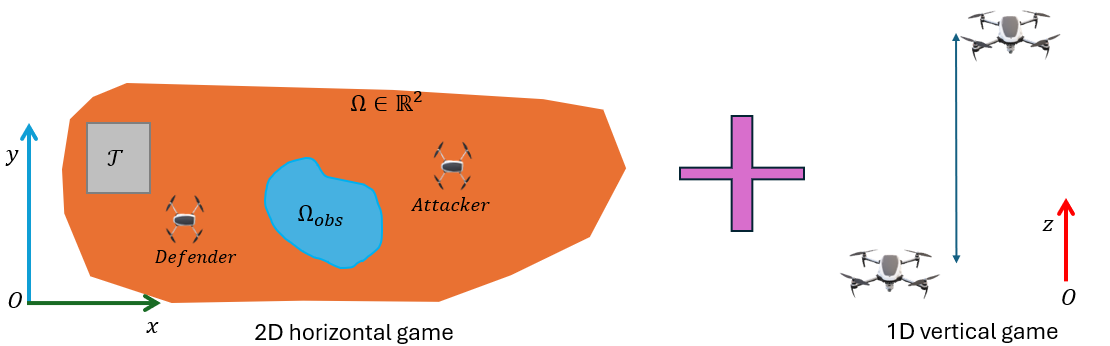}
    \caption{Our approach breaks down the problem into horizontal and vertical game}
    \label{fig:decompose_figure}
\end{figure}

\subsubsection{Horizontal Game}

In the horizontal game, we will only consider the horizontal $x$-$y$ components in the full dynamics:

\begin{equation} \label{2d_dynamics} 
        \dot{\mathbf{x}}_{D}^{h} = 
        \begin{bmatrix} 
            \dot{x}_{D} \\ \dot{y}_{D} \\ \dot{v}^{x}_D \\ \dot{v}^{y}_D 
        \end{bmatrix} 
        = 
        \begin{bmatrix} v^{x}_D\\ v^{y}_D \\ k_x(v^{x, C}_{D} - v^{x}_D)\\ k_y(v^{y, C}_{D} - v^{y}_D) 
        \end{bmatrix}, \qquad
        \dot{\mathbf{x}}_{A}^{h} = 
        \begin{bmatrix} 
            \dot{x}_{A} \\ \dot{y}_{A} 
        \end{bmatrix} 
        = 
        \begin{bmatrix} 
            v^{x, C}_{A} \\ v^{y, C}_{A} 
        \end{bmatrix} 
\end{equation} 

We will shortly denote the joint state in the horizontal direction
$\mathbf{x}^{h} = (\mathbf{x}^{h}_{D}, \mathbf{x}^{h}_{A}) \in \mathbb{R}^6 $  and the control for each attacker and defender as
$\mathbf{u}^{h}_{D} = \left(v^{x, C}_D, v^{y, C}_D\right) \in \mathbb{R}^2, \mathbf{u}^{h}_{A} = \left(v^{x, C}_A, v^{y, C}_A\right) \in \mathbb{R}^2$.
Similar to the original reach-avoid game formulation, the reach and avoid set in the horizontal game are defined as follows:

\begin{subequations}
    \label{eq:origin_horizontal_reach_avoid_set} 
    \begin{align} 
        \mathcal{R}_{h} &=  \left\{\mathbf{x}^{h} \mid \mathbf{x}^{h}_{A} \in \mathcal{T}  \right\} \cup \left\{ \mathbf{x}^{h} \mid \mathbf{x}^{h}_{D} \in \Omega_{o b s}\right\} \label{eq:original_horizontal_reach_set}\\ 
        \mathcal{A}_{h} &= \left\{ \mathbf{x}^{h} \mid \left\|\mathbf{x}^{h}_{A}- \mathbf{x}^{h}_D\right\|_2 \leq d_{h} \wedge \mathbf{x}^{h}_{A} \notin \mathcal{T} \right\} 
        \cup\left\{\mathbf{x}^{h} \mid \mathbf{x}^{h}_{A} \in \Omega_{o b s}\right\} \label{eq:original_horizontal_avoid_set}
    \end{align} 
\end{subequations} 

Since the dynamics of the defender and attacker are decoupled, the Hamiltonian value in Eq.~\eqref{eq:Hamiltonian} is independent of the min-max order, and hence
the value function of the horizontal game  $\Phi_{h}(\mathbf{x}^{h}_{A}, \mathbf{x}^{h}_{D})$ is well-defined and can be obtained by solving Eq.~\eqref{eq:HJI}.
The joint state $\mathbf{x}^{h}$ is a 6D vector, and numerically solving Eq.~\eqref{eq:HJI} for the joint state is at the limit of tractable computation using software toolbox such as \cite{optimized}.
Optimal feedback controls in the horizontal case can be computed using spatial derivatives:

\begin{subequations}
    \label{eq:opt_reach_horizontal_ctrl}  
     \begin{align}
        &(\mathbf{u}^{h}_{D})^{*}(\mathbf{x}, t)=\arg \max _{\mathbf{u}^{h}_{D} } \left(\frac{\partial \Phi_{h}}{\partial \mathbf{x}^h} \right)^{\top} f_{h}(\mathbf{x}^{h}, \mathbf{u}^{h}_{A}, \mathbf{u}^{h}_{D}) \label{eq:opt_reach_horizontal_ctrl_defender}  \\
        &(\mathbf{u}^{h}_{A})^{*}(\mathbf{x}, t)=\arg \min _{\mathbf{u}^{h}_A} \left(\frac{\partial \Phi_{h}}{\partial \mathbf{x}^{h}} \right) ^{\top}f_{h}\left(\mathbf{x}^{h}, \mathbf{u}^h_{A}, (\mathbf{u}_{D}^{h})^{*} \right) \label{eq:opt_reach_horizontal_ctrl_attacker} 
     \end{align}
\end{subequations}
where $f_{h} = \begin{bmatrix} \dot{\mathbf{x}}^{h}_{D} \\  \dot{\mathbf{x}}^{h}_{A} \end{bmatrix}$ is the joint dynamics in horizontal plane.
As before, the winning region for the attacker and defender in horizontal case are the sub-zero and super-zero level set of $\Phi_{h}$, respectively:

\begin{subequations} \label{eq:original_horizontal_winning_region}
    \begin{align}
        \mathcal{W}_{A, h} &= \{\mathbf{x}^{h} \mid \Phi_{h}(\mathbf{x}^{h }) \leq 0\} \label{eq:original_horizontal_winning_region_a} \\
        \mathcal{W}_{D, h} &= \{\mathbf{x}^{h} \mid \Phi_{h}(\mathbf{x}^{h }) > 0\} \label{eq:original_horizontal_winning_region_d}
    \end{align}
\end{subequations}

\subsubsection{Vertical Game}

In the vertical game, the dynamics involve only the vertical components of the defender and attacker:

\begin{equation} \label{1d_dynamics} 
    \begin{aligned} 
        \dot{\mathbf{x}}_{{D}}^{z} = \begin{bmatrix} \dot{z}_{D} \\ \dot{v}^{z}_D \end{bmatrix} &= \begin{bmatrix} v^{z}_D \\\ k_z(v^{z, C}_{D} - v^{z}_D) \end{bmatrix}, & \dot{\mathbf{x}}_{{A}}^{z} = \begin{bmatrix} \dot{z}_{A} \\ \end{bmatrix} &= \begin{bmatrix} v^{z, C}_{A} \end{bmatrix} 
    \end{aligned} \end{equation} 

We will shortly denote the joint state in the horizontal direction $\mathbf{x}^{z} = (\mathbf{x}_{D}^{z}, \mathbf{x}_{A}^{z}) \in \mathbb{R}^3 $  and the control for each attacker and defender as $\mathbf{u}^{z}_{D} = v^{z, C}_D, \mathbf{u}^{z}_{A} = v^{z, C}_A $.

Since the goal region and obstacle set are not dependent on the vertical height, we can define the reach set (for the defender) as follows:

\begin{equation} \begin{aligned}  \mathcal{R}_{z} &= \left\{ \mathbf{x}^{z} \mid \abs{z_A -z_D} \leq d_{z} \right\} \end{aligned} \end{equation} 

Thus, we only need to solve for a simpler variational inequality without obstacle function $g(x)$:

\begin{equation} \label{pde: HJI2} \begin{aligned} \min \left\{\frac{\partial \Phi_z}{\partial t} + H \left(\mathbf{x}^z, \frac{\partial \Phi_z}{\partial \mathbf{x}^z}\right), l_{z}(\mathbf{x}^z)-\Phi_z(\mathbf{x}^z, t) \right\} &=0, \quad t \in[0, T] \\ \Phi_z(\mathbf{x}^z,  T ) &= l_{z}(\mathbf{x}^z) \end{aligned} \end{equation}
where $l_z = \abs{z_A -z_D} - d_z$.

In this game, the defender will try to minimize the relative distance function $\Phi_z(\mathbf{x}_z, t )$ while the attacker will maximize.
These objectives are reversed compared to the horizontal sub-game 
and the original formulation in Eq. \eqref{eq:game_value}, where defender minimizes while attacker maximizes.
This is because the objective function being considered is the distance between attacker and defender, which is implicit in our choice of the reach set 
for the defender.
We solve the above equation to obtain the value function $\Phi_{z}(\mathbf{x}^z_{D}, \mathbf{x}^z_{A})$ from which optimal controls for both agents are computed: 

\begin{subequations} \label{eq:opt_reach_vertical_ctrl} 
    \begin{align} 
        &  (\mathbf{u}^{z}_{D})^{*}(\mathbf{x})=\arg \min _{\mathbf{u}^z_{D} } \left(\frac{\partial \Phi_{z}}{\partial \mathbf{x}^{z}} \right) ^{\top} f_{z}(\mathbf{x}^{z}, \mathbf{u}_{A}^z, \mathbf{u}_{D}^z) \\
        &  (\mathbf{u}^{z}_{A})^{*}(\mathbf{x})=\arg \max _{\mathbf{u}^z_{A}} \left(\frac{\partial \Phi_{z}}{\partial \mathbf{x}^{z}} \right)^{\top} f_{z}\left(\mathbf{x}^{z}, \mathbf{u}^z_{A}, (\mathbf{u}^z_{D})^{*}\right) 
    \end{align} 
\end{subequations} 

\noindent where $f_{z} = \begin{bmatrix} \dot{\mathbf{x}}^{z}_{D} \\  \dot{\mathbf{x}}^{z}_{A} \end{bmatrix}$ is the system dynamics in the vertical direction.
In addition, the winning region in vertical game for the attacker and defender can be obtained using the level set of this function, respectively:

\begin{subequations} \label{eq:original_vertical_winning_region}
    \begin{align}
        \mathcal{W}_{D, z} = \{\mathbf{x}^{z} \mid \Phi_{z}(\mathbf{x}^{z}) \leq 0\}  \label{eq:original_vertical_winning_region_d} \\
        \mathcal{W}_{A, z} = \{\mathbf{x}^{z} \mid \Phi_{z}(\mathbf{x}^{z}) > 0\}  \label{eq:original_vertical_winning_region_a}
    \end{align}
\end{subequations}

For all $\mathbf{x} \in \mathcal{W}_{D, z}$ we also define the earliest time the defender captures the attacker vertically when each agent applies its optimal control function:

\begin{equation} \label{eq:time_to_capture} T_{\text{capture}}(\mathbf{x}^{z}) = \min \left\{t \mid  \zeta^{(\mathbf{u}^{z}_{A})^{*}(\cdot),  (\mathbf{u}^{z}_{D})^{*}(\cdot)}_{\mathbf{x}^{z}, 0}(t) \in \mathcal{R}_{z} \right\}, \end{equation} which can be numerically extracted by considering the earliest time where $\Phi_z(\mathbf{x}^z,  t )$ change its sign from positive to negative or zero during the numerical process of solving Eq.
(\ref{pde: HJI2}).

\section{Invariant Capturing Control Set}

From the result of previous sections, one can attempt to apply the optimal controls computed from each sub-game for the defender 
while it is inside the winning region of each sub-game. The problem of doing this is we cannot guarantee if the defender 
captures the attacker in the horizontal game at the same time as in the vertical game.
Since the defender is a double-integrator dynamics
in every direction, the defender can arrive and capture the attacker at a velocity that is bound to move out of the capture set
because of inertia, or even worse, out of the winning region thereafter.
Hence, the capture conditions in \eqref{eq:horizontal_condition} and \eqref{eq:vertical_condition} are not guaranteed to be satisfied simultaneously if we just
simply combine the sub-games' results.

In the following sections, our aim is to improve this situation by considering invariant capture sets in each subgames.
This property will allow us to overcome the difficulty aforementioned and better analyze winning conditions when combining
the winning regions of each sub-game, as will be shown in the section V and VI.

\subsection{Invariant Vertical Capture Set}

Consider relative dynamics of the defender and attacker in the vertical sub-game: 

\begin{equation} 
    \dot{\mathbf{x}}_{z}^\text{rel} =  f^\text{rel}_{z}(\mathbf{x}^\text{rel}_{z}, \mathbf{u}^{z}_{D}, \mathbf{u}^{z}_{A}) = 
    \begin{bmatrix} \dot{z}^\text{rel} \\ \dot{v}^{z}_D \end{bmatrix} 
    = \begin{bmatrix} v^z_D - v^{z, C}_A \\ k_z(v^{z, C}_{D} - v^{z}_D) \end{bmatrix}
\end{equation} 

\noindent where $z^\text{rel} = z_D -z_A$.

Define the distance function $l_z(z_{rel}) = \abs{z_{rel}}$ for the vertical axis., and consider the maximum relative vertical distance over time under optimal controls by both the defender and attacker in the horizontal game: 

\begin{equation} \label{eq:V_z} 
    V_{z}(\mathbf{x}^\text{rel}_{z}, t) = \sup_{\mathbf{u}^z_A (\cdot)} \inf_{\mathbf{u}^z_D (\cdot)}  \max_{\tau \in [0, t]} l_z(\zeta^{\mathbf{u}^z_D(\cdot), \mathbf{u}^z_A(\cdot)}_{\mathbf{x}^\text{rel}_{z}, 0}(\tau)) 
\end{equation} 

Note that this value considers the maximum relative distance driven by the optimal defender and adversarial attacker from the beginning to the current time $t$. This is different from functional value in Eq. \eqref{eq:V_z_t},
which considers the minimum value from current time $t$ to the terminal time $T$. In this function, the defender aiming to track the attacker minimizes the relative distance
while the attacker maximizes. These optimizing roles are different from those in Eq. \eqref{eq:V_z_t} as the objective functions are different in both cases. Intuitively, this maximum relative distance is the upper bound of the relative distance given the initial joint state over time. The significance of the function $V$ is,
no matter what the attacker chooses to do, there exists a defender's control function that can keep their relative
distance at most $V$ over time. The above value function could be solved
using the formulation in \cite{FastTrackV2} and  obtained using reachability toolbox such as toolboxLS or
optimizedDP \cite{optimized}.
We denote the value function converging at infinite horizon time as $V_{z, \infty}(\mathbf{x}^\text{rel}_z) = \lim_{t \rightarrow \infty}
V(\mathbf{x}^\text{rel}_z, t)$.
Next, consider the following level-set where capture in z-axis satisfies: $\mathcal{B}_{z}= \{\mathbf{x}^\text{rel}_{z} \mid V_{z, \infty}(\mathbf{x}^\text{rel}_{z}) \leq d_z \}$.
For now, we assume this level-set exists if the defender's control is powerful enough.
Alternatively, it is also possible to set the captured distance $d_z$ to be the minimum level set of $V_{z, \infty}$.

\begin{remark}
    \label{remark: vertical_invariant}
    \textit{The set $\mathcal{B}_{z} = \{\mathbf{x}^\text{rel}_{z} \mid V_{z, \infty}(\mathbf{x}^\text{rel}_{z}) \leq d_z \}$ is a control invariant set.
        In other words, ,  if $\mathbf{x}^\text{rel}_{z}(t_1) \in \mathcal{B}_{z} \implies \zeta^{(\mathbf{u}^z_D)^{*}(\cdot), 
        (\mathbf{u}^z_A)^{*}(\cdot)}_{\mathbf{x}^\text{rel}_z, t_{1}}(t_2) \in \mathcal{B}_{z}, \space \forall t_2 \ge t_1$ where} 

    \begin{equation} \label{eq:opt_d_track_u} (\mathbf{u}^z_D)^{*}(\cdot) = \arg \inf_{\mathbf{u}^z_D(\cdot)}\max_{\tau \in [t_1, \infty]} l_z\left(\zeta^{\mathbf{u}^z_D(\cdot),\mathbf{u}^z_A(\cdot)}_{\mathbf{x}^\text{rel}_z, t_1}(\tau)\right), \end{equation} \begin{equation} \label{eq:opt_a_track_d} (\mathbf{u}^z_A)^{*}(\cdot) = \arg \sup_{\mathbf{u}^z_{A}(\cdot)} \inf_{\mathbf{u}^z_D(\cdot) }\max_{\tau \in [t_1, \infty]} l_z\left(\zeta^{\mathbf{u}^z_D(\cdot),\mathbf{u}^z_A(\cdot)}_{\mathbf{x}^\text{rel}_z, t_1}(\tau)\right), \end{equation} \end{remark} 

In fact, the above remark holds true for all level set of $V_{z, \infty}$, which has been proven in \cite{FastTrack}.
The above result implies that once $ \mathbf{x}^\text{rel}_{z}(t_1) \in \mathcal{B}_z$, the defender is, indefinitely, guaranteed to keep a distance at most $d_z$, and hence maintain the captured status (in the vertical direction) if it applies optimal control in Eq.~\eqref{eq:opt_d_track_u}.
If deep inside the set, defender can switch to use another performant controller that is less aggressive to follow the attacker, such as a PID controller.
On the other hand, if the joint system state is on the boundary of $\mathcal{B}_{z}$, the defender is guaranteed to stay inside the set using the optimal tracking controller, according to the gradient of $V_{z, \infty}$: 

\begin{equation} \label{eq:opt_tracking_vertical} 
    (\mathbf{u}^{z}_D)^{*} = \arg \min_{ \mathbf{u}^{z}_D}\max_{\mathbf{u}^{z}_{A} } \left(\dfrac{\partial V_{z, \infty}}{\partial \mathbf{x}^\text{rel}_{z}} \right)^{\top} f^\text{rel}_{z}(\mathbf{x}^\text{rel}_{z}, \mathbf{u}^{z}_{D}, \mathbf{u}^{z}_{A}) 
\end{equation} 

We will call the controller obtained above a vertical \textbf{tracking} controller.
If the attacker acts sup-optimally and the defender acts optimally, the joint system state can drop a lower level set and the vertical distance can even be smaller.

\subsection{Invariant Horizontal Captured Set}

For the horizontal game, we will denote the relative distance function $l^{\text{rel}}_{h}(\mathbf{x}_h) = \sqrt{(x^{\text{rel}})^2 +(y^{\text{rel}})^2}$ where $x^\text{rel} = x_D - x_A$, and $y^\text{rel} = y_D - y_A$.
Unlike from the vertical sub-game, there are obstacles in the horizontal sub-game.
If we only consider the relative distances as in previously
\begin{equation} \label{eq:V_h_rel_only} 
    V_{h}(\mathbf{x}^\text{rel}_{h}, t) = \sup_{\mathbf{u}^h_A (\cdot)} \inf_{\mathbf{u}^h_D (\cdot)}  \max_{\tau \in [0, t]} l_h^{\text{rel}}(\zeta^{\mathbf{u}^h_D(\cdot), \mathbf{u}^h_A(\cdot)}_{\mathbf{x}^\text{rel}_{h}, 0}(\tau)) 
\end{equation}
the invariant tracking set does not take
into account obstacle avoidance for the defender. In order to ensure the defender avoids obstacles while maintaining the captured status
for the defender,
we alternatively consider the following target function that accounts for both relative distance and static obstacles in the horizontal game:
\begin{equation} \label{eq:horizontal_value_func_obs}
\begin{aligned}
     V_{h}(\mathbf{x}_h, t) = \sup_{\mathbf{u}^h_A (\cdot)} \inf_{\mathbf{u}^h_D (\cdot)} \max_{\tau \in [0, t]} \min \left\{l_h\left(\zeta^{\mathbf{u}^h_D (\cdot),
    \mathbf{u}^h_A(\cdot)}_{\mathbf{x}_{h}, 0}(\tau)\right), \min_{s \in [0, \tau]} o_{h, A}\left(\zeta^{\mathbf{u}^h_D (\cdot), \mathbf{u}^h_A(\cdot)}_{\mathbf{x}_{h}, 0}(s)\right)\right\} 
\end{aligned} 
\footnote{In the published journal version, the function is $V_{h}(\mathbf{x}_h, t) = \sup_{\mathbf{u}^h_A (\cdot)} \inf_{\mathbf{u}^h_D (\cdot)} \max_{\tau \in [0, t]} l_h(\mathbf{x}_h, t)$ . However, this value can be conservative as it does not force the attacker to avoid obstacles.}
\end{equation} 
where $l_h(\mathbf{x}) = \max\left\{l^{\text{rel}}_{h}(\mathbf{x}), o_{h,D}(\mathbf{x})\right\}$ with $o_{h,D}(\mathbf{x}) =     \begin{cases}
    K, & \text{if } (x_D, y_D) \in \Omega_{\text{obs}} \\
    0, & \text{else }
\end{cases}$, $o_{h, A}(\mathbf{x}) = \begin{cases}
    0, & \text{if } (x_A, y_A) \in \Omega_{\text{obs}} \\
    K, & \text{else }
\end{cases}$ and $K$ is some large positive constant. Notice that this value function is very similar to the equation
\eqref{eq:V_z_t} where the max is swapped with min and vice versa, which can be obtained via solving
the pde in \eqref{eq:HJI} with max and min operation swapped. In the absence of obstacle,
above equation just reduces to eq. \eqref{eq:V_h_rel_only}, similar to the vertical case.
The function $l_h(\mathbf{x}) = \max\left\{l^{\text{rel}}_{h}(\mathbf{x}), o_{h,D}(\mathbf{x})\right\}$ captures the fact that defender tracking the attacker while avoiding obstacle.
Intuitively, a defender colliding with an obstacle would be considered to have lost track of the attacker, which is equivalent to the tracking distance $l_h$ being very large K. 

On the attacker side, if it hits the 
obstacle at any time, the momentary tracking distance $o_{h, A}$ would be zero,
meaning the defender wins immediately. The innermost term $\min_{s \in [0, \tau]} o_{h, A}\left(\zeta^{\mathbf{u}^h_D (\cdot), \mathbf{u}^h_A(\cdot)}_{\mathbf{x}_{h}, 0}(s)\right)$ checks whether the attacker
hits obstacle at any time $s$, if it has not hit obstacle before, the outer min would just resolve taking minimum of
$l_h$ and $K$ and hence just becomes $l_h$. However if attacker hits obstacle at any point, 
$o_{h,A}$ would return zero and the outer min would just return zero. Since the outer max takes the 
maximum of the inner value over time, if attacker hits obstacle, the
maximum tracking distance does not increase thereafter since the defender has won the game at that point.  

By minimizing the objective value, the defender will try to both
minimize distance to the attacker and the cost function of obstacles over time, or pushing the attacker to collide with
obstacles. Similar to the vertical case,
this value function could be solved using the formulation in \cite{FastTrackV2} and obtained using reachability toolbox such as toolboxLS or optimizedDP \cite{optimized}. 
We also consider captured level-set of the above function: $\mathcal{B}_{h} = \{\mathbf{x}_h \mid V_{h, \infty}(\mathbf{x}_{h}) \leq d_{h} \}$.
Similar to the vertical case, we have the following:

\begin{remark} \label{remark: horizontal_invariant} The set $\mathcal{B}_{h} = \{\mathbf{x}_{h} \mid V_{h, \infty}(\mathbf{x}_{h}) \leq d_{h} \}$ is a control invariant set.
    In other words, ,  if $\mathbf{x}_{h}(t_1) \in \mathcal{B}_{h} \implies \zeta^{(\mathbf{u}^{h}_D)^{*}(\cdot), (\mathbf{u}^{h}_A)^{*}(\cdot)}_{\mathbf{x}_{h}, t_1}(t_2) \in \mathcal{B}_{h}, \space \forall t_2 \ge t_1$ where 

    \begin{equation} \label{eq:opt_d_track_horizontal} (\mathbf{u}^{h}_D)^{*}(\cdot) = \arg \inf_{\mathbf{u}^h_D(\cdot)}\max_{\tau \in [t_1, \infty]} \min \left\{l_h\left(\zeta^{\mathbf{u}^h_D (\cdot),
    \mathbf{u}^h_A(\cdot)}_{\mathbf{x}_{h}, t_1}(\tau)\right), \min_{s \in [t_1, \tau]} o_{h, A}\left(\zeta^{\mathbf{u}^h_D (\cdot), \mathbf{u}^h_A(\cdot)}_{\mathbf{x}_{h}, t_1}(s)\right)\right\}, \end{equation} \begin{equation} \label{eq:opt_a_track_horizontal} (\mathbf{u}^{h}_A)^{*}(\cdot) = \arg \sup_{\mathbf{u}^h_{A}(\cdot)} \inf_{\mathbf{u}^h_D(\cdot) }\max_{\tau \in [t_1, \infty]} \min \left\{l_h\left(\zeta^{\mathbf{u}^h_D (\cdot),
    \mathbf{u}^h_A(\cdot)}_{\mathbf{x}_{h}, t_1}(\tau)\right), \min_{s \in [t_1, \tau]} o_{h, A}\left(\zeta^{\mathbf{u}^h_D (\cdot), \mathbf{u}^h_A(\cdot)}_{\mathbf{x}_{h}, t_1}(s)\right)\right\}, \end{equation} \end{remark} 

If $ \mathbf{x}_{h}(t_1) \in \mathcal{B}_{h}$, meaning that $V_{h, \infty}(\mathbf{x}_{h}(t_1)) \leq d_{h} < K$, then the defender is, indefinitely, guaranteed to keep a distance at most $d_{h}$ and maintain the captured status 
while avoiding obstacles if it applies optimal control in Eq. \eqref{eq:opt_d_track_horizontal}.
Once inside the set, defender can apply a performant controller to follow the attacker such as tracking attacker's velocity and positions.
The defender is guaranteed to stay inside the set with a \textbf{horizontal} tracking controller computed using $V_{h, \infty}$:
\begin{equation}
    \label{eq:opt_control_track_horizontal}
    (\mathbf{u}^{h}_D)^{*} = \arg \min_{ \mathbf{u}^{h}_D}\max_{\mathbf{u}^{h}_{A}
    } \left(\dfrac{\partial V_{h, \infty}}{\partial \mathbf{x}_{h}} \right)^{\top} f_{h}(\mathbf{x}_{h}, \mathbf{u}^{h}_{D}, \mathbf{u}^{h}_{A})
\end{equation}

\section{A Reach-Track control algorithm}

\begin{figure}
    \centering
    \includegraphics[scale=0.62]{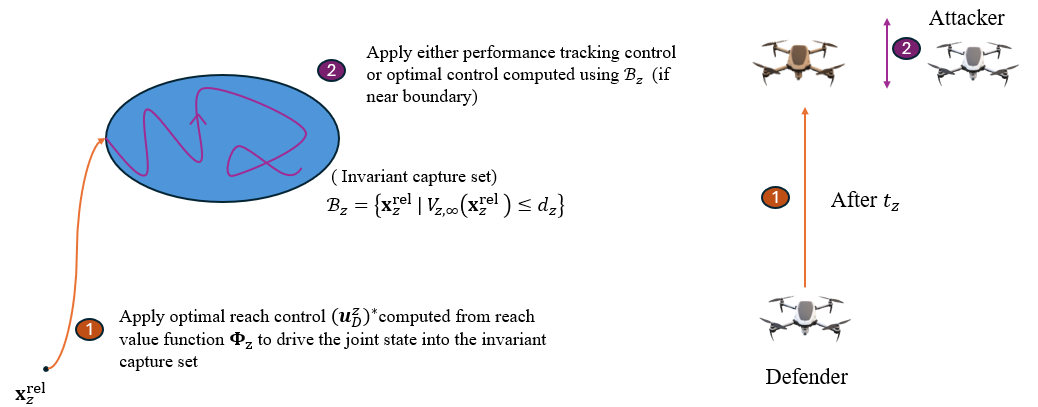}
    \caption{Our Reach-Track control procedure in vertical axis}
    \label{fig:invariant_set}
\end{figure}
\subsection{Vertical Reach-track}

Back to the vertical sub-game, we will modify its reach-set as follows: 

\begin{equation} \label{eq:new_A_z} 
    \mathcal{R}_z = \left\{\mathbf{x}^{z} \mid \mathbf{x}^{rel}_{z} \in \mathcal{B}_{z} \right\},
\end{equation}

\noindent where we replace the standard captured set $\mathcal{C}_z = \{ \mathbf{x}^{z} \mid \sqrt{(z_A -z_D)^2 } \leq d_z \}$ with the invariant capture set $\mathcal{B}_{z}$ introduced previously.
In general, $ \mathcal{B} _z\subseteq \mathcal{C}_z $, meaning $\mathcal{B}_z$ is more restricted than $\mathcal{C}_{z}$, 
which we will numerically validate in later experiment section.
This makes sense physically; for example, if the defender needs to fly up or down to match the attacker's height and maintain the difference thereafter, it needs to arrive at approximately the same speed as the attacker without overshooting and disrupting the capture status.
Once inside the new reach set, the defender will not only capture the attacker vertically but also be guaranteed to maintain the capture status if applying the corresponding state-feedback control.

We then solve Eq.~\eqref{pde: HJI2} with $l_z(\mathbf{x}^z) = V_{z, \infty}(z^{rel}) - d_z$ now being the level-set representation of $\mathcal{B}_z$ to obtain a new game value function $\Phi_z$ and winning region $\mathcal{W}_{D, z}$.
Our vertical capture procedure now consists of two stages.
The first stage is to apply the optimal control computed using  $\Phi_z$ obtained with reach set $\mathcal{R}_z$ defined in eq.
(\ref{eq:new_A_z}) to arrive in $\mathcal{B}_{z}$ and capture attacker in the shortest time.
Once the joint state arrives in $\mathcal{B}_z$, the defender will switch to use the optimal controller derived from function $V_{z, \infty}(\mathbf{x}^{rel}_{z})$ to track the attacker.
If the joint state is deep inside the invariant capture set, a performant PID controller can be used instead to track the attacker's vertical position and velocity.
This procedure is described in Algorithm \ref{alg:vertical_reach_track} and illustrated in Figure \ref{fig:invariant_set}.

\begin{algorithm} \caption{Vertical Reach-Track control algorithm}\label{alg:vertical_reach_track} \begin{algorithmic}[1] \While{$\mathbf{x}_{z} \in \mathcal{W}_{D, z}$} \Comment{While joint vertical state is in winning barrier}: \If{$\mathbf{x}^{rel}_{z} \notin \mathcal{B}_{z}$, or $ V_{z, \infty}(\mathbf{x}^{rel}_z) > d_z$ } \Comment{Attacker is not yet captured vertically} \State Apply $(\mathbf{u}^{z}_D)^{*}$ using Eq.
        \ref{eq:opt_reach_vertical_ctrl} \Comment{Apply optimal control to reach $\mathcal{B}_{z}$ in shortest time}
        \Else
        \If{$V_{z, \infty}(\mathbf{x}^{rel}_z) - d_z \leq \epsilon$ } \Comment{Joint state is deep inside invariant capture set}
        \State Apply any performant tracking controller for vertical control
        \Else \Comment{Near the boundary of $\mathcal{B}_{z}$}
        \State Apply $(\mathbf{u}^{z}_D)^{*}$ using Eq. \ref{eq:opt_tracking_vertical} \Comment{Apply optimal control to stay inside $\mathcal{B}_z$}
        \EndIf
        \EndIf
        \EndWhile

    \end{algorithmic}
\end{algorithm}
The combination of reachability and invariant set basically recovers a stabilizing Lyapunov function, however analytically deriving Lyapunov function can be challenging.
Hence, the numerical computation of invariant set first and then applying the reachability toolbox may provide a more general procedure.

\subsection{Horizontal Reach-Track-Avoid algorithm }

In the vertical reach-avoid game, similar to before, we replace the horizontal capture set with the invariant capture set $\mathcal{B}_{h}$.
The reach set and avoid set for horizontal are as follows: 

\begin{equation} \begin{aligned} \mathcal{R}_{h} &=  \left\{\mathbf{x}_{h} \mid \mathbf{x}^{h}_{\mathbf{A}} \in \mathcal{T}\right\} \cup \left\{ \mathbf{x}^{h} \mid \mathbf{p}^{h}_{D} \in \Omega_{o b s}\right\} \\ \mathcal{A}_{h} &= \left\{ \mathbf{x}_{h} \mid \mathbf{x}^{rel}_{h} \in \mathcal{B}_{h} \wedge \mathbf{x}^{h}_{\mathbf{A}} \notin \mathcal{T} \right\} \cup\left\{\mathbf{x} \mid \mathbf{x}_{A} \in \Omega_{o b s}\right\} \end{aligned} \end{equation} 

\begin{algorithm} \caption{Horizontal Reach-Avoid-Track control algorithm for defender}\label{alg:horizontal_reach_avoid_track} \begin{algorithmic}[1] \While{$\mathbf{x}_{h} \in \mathcal{W}_{D,h}$} \Comment{While joint horizontal state is still in defender's winning region} \If{$\mathbf{x}^{rel}_{h} \notin \mathcal{B}_{h}$} \Comment{Apply optimal control to reach $\mathcal{B}_{h}$ and capture the attacker} \State Apply $(\mathbf{u}^{h})^{*}_{D}$ using Eq.
        (\ref{eq:opt_reach_horizontal_ctrl})
        \Else \Comment{The defender already captures the attacker horizontally and inside $\mathcal{B}_{h}$}
        \If{$V_{h, \infty}(\mathbf{x}^{rel}_{h}) - d_{h} \leq \epsilon_{track}$} \Comment{Joint state is deep inside $\mathcal{B}_{h}$}
        \State Apply performant controller for $\mathbf{u}^{h*}_{D}$
        \Else
        \State Apply $(\mathbf{u}^{h})^{*}_{D}$ using Eq. (\ref{eq:opt_control_track_horizontal}) \Comment{Apply optimal control to maintain captured status}
        \EndIf
        \EndIf
        \EndWhile

    \end{algorithmic}
\end{algorithm}

We then solve Eq.~\eqref{pde: HJI2} for $\Phi_h$ and obtain a new winning region for each player.
The capturing procedures for the defender are shown in Algorithm \ref{alg:horizontal_reach_avoid_track}.
Different from the vertical case, there are obstacles in the scene the defender needs to avoid.
Since our computed invariant capture set $\mathcal{B}_{h}$ already takes into account the obstacles during the tracking
process, once the joint state is inside $\mathcal{B}_{h}$, the defender can apply the optimal tracking control derived
from $V_{h, \infty}$ to maintain the capture status while avoiding obstacles (line 11 of Algorithm \ref{alg:horizontal_reach_avoid_track}).

Alternatively, if $V_{h, \infty}$ does not account for obstacle during tracking,
one need to have a switching safety controller and compute a control
that aims to satisfy both tracking and obstacle avoidance inequality constraints of a quadratic optimization program as in \cite{QP} \cite{CBFReview}.
However, the interference of the safety filter might lead to sub-optimal control that breaks the
invariant capture guarantee. 

By applying both Algorithm 1 and 2, we can keep the game analysis tractable while avoiding the intractable direct computation or 
the back-projection operation \cite{DecompositionMethod} in the 9D state space. Since the invariant tracking set is more conservative and smaller in  than the original capture set, the new winning region for 
the defender can also be smaller than the original winning region obtained in section \textbf{III}.C, especially in the
horizontal game. To mitigate this conservative property, instead of applying both Algorithms \ref{alg:vertical_reach_track} and \ref{alg:horizontal_reach_avoid_track} simultaneously,
one can alternatively consider only tracking the attacker in the "easier" sub-game while using original capture strategy for the other sub-game.
For example, we can consider applying Algorithm 1 to the vertical sub-game while only using the original capture strategy to collide with the attacker
in the horizontal game. 
More detailed treatment of this strategy will be considered in the future work. 
For the rest of the paper, we will focus on the game result when applying both Algorithms \ref{alg:vertical_reach_track} and \ref{alg:horizontal_reach_avoid_track} on the defender.

\section{Capture Guarantee Analysis}
In this section, we will analyze the conditions for the defender to win the original 9D game when applying Algorithms \ref{alg:vertical_reach_track} and \ref{alg:horizontal_reach_avoid_track}.
Before then, let's consider optimal reaching for the attacker to the target region \textbf{without} taking avoiding the defender into account.
This would require solving Eq.
\eqref{eq:HJI} but with a simpler avoid set and reach set in the horizontal direction:
\begin{equation}
    \begin{aligned}
        \mathcal{R}_{h} & =  \left\{\mathbf{x}^{h}_{A} \mid \mathbf{x}^{h}_{A} \in \mathcal{T}\right\} \\
        \mathcal{A}_{h} & =\left\{\mathbf{x}^{h}_{A} \mid \mathbf{x}^h_{A} \in \Omega_{o b s}\right\}
    \end{aligned}
\end{equation}

We will then solve for the corresponding value function $\Phi_{A, h}^{\text{reach}}$, for which we can extract
\begin{equation}
    T_{\text{goal}}(\mathbf{x}) = \min \{ t \mid  \zeta^{\mathbf{u^{*}_{A}}(\cdot)}_{\mathbf{x}}(t) \in \mathcal{T} \} 
\end{equation} 

\noindent as the shortest time  for the attacker to reach goal by applying the optimal reaching control 
\begin{equation} \label{eq:optimal_reaching_to_goal} 
    \mathbf{u}_{A}^{*}(\mathbf{x}, t)=\arg \min _{\mathbf{u}_A} \frac{\partial \Phi_{A, h}^{\text{reach}}}{\partial \mathbf{x}} f_{A, h}(\mathbf{x}, \mathbf{u}_{A}).
\end{equation} 

For states where the attacker inevitably run into the obstacles, their time to goal are infinitely large.

Next, suppose that $\mathbf{x}_{h} \in \mathcal{W}_{D, h}$ (defined in Eq.~\eqref{eq:original_horizontal_winning_region_d}) and $\mathbf{x_{z}} \in \mathcal{W}_{D, z}$ (Eq.~\eqref{eq:original_vertical_winning_region_d}) computed with the invariant sets and the defender applies Algorithm \ref{alg:vertical_reach_track} and \ref{alg:horizontal_reach_avoid_track}.
If these conditions are satisfied, it is still not the case that the defender is guaranteed to capture the attacker in the original 9D game.
In fact, we will need an extra condition to determine if the attacker or defender can win the original game.

\textit{\textbf{Proposition 1:}
    If $T_{\text{goal}}(\mathbf{x^{h}_{A}}) \leq T_\text{capture}(\mathbf{x^{z}})$ (defined in Eq. \eqref{eq:time_to_capture})
     the attacker can win the game.
}

\textit{Proof:}
Horizontally, consider that the attacker simply follows the shortest path by applying control in Eq.~\eqref{eq:optimal_reaching_to_goal} to reach the goal in shortest time while avoiding obstacles, which is $T_{\text{goal}}(\mathbf{x^{h}_{A}})$.
On vertical axis, it applies optimal control in Eq.~\eqref{eq:opt_reach_vertical_ctrl} to achieve maximum capture time of  $T_\text{capture}(\mathbf{x^{z}})$.
If $T_{\text{goal}}(\mathbf{x^{h}_{A}}) \leq T_\text{capture}(\mathbf{x^{z}})$ the attacker will arrive at the goal region before capture happens in the vertical axis, and hence win the game no matter if the attacker has been captured in the horizontal game or not.

The above condition is guaranteed for the attacker to win the game when the defender applies Algorithms \ref{alg:vertical_reach_track} and \ref{alg:horizontal_reach_avoid_track}.
On the other hand, let's consider the scenarios when the defender is guaranteed to win the game using Algorithms \ref{alg:vertical_reach_track} and \ref{alg:horizontal_reach_avoid_track}.

\textbf{Proposition 2:}  \textit{If  $\mathbf{x_{h}} \in \mathcal{W}_{D, h}$ and $\mathbf{x}^{rel}_{z} \in \mathcal{B}_z$ the defender is guaranteed to win.}

\textit{Proof: }
Since $\mathbf{x_{h}} \in \mathcal{W}_{D, h}$, the defender is guaranteed to capture the attacker and wins the horizontal game at some time in the future using optimal control in Eq.
(\ref{eq:opt_reach_horizontal_ctrl}),  by the definition of the winning regions.
Up until then, because $\mathbf{x}^{rel}_{z} \in \mathcal{B}_z$, it must be true that $|\mathbf{p}^{z}_A - \mathbf{p}^{z}_D| \leq d_z$ at all times if it applies optimal tracking control Algorithm \ref{alg:vertical_reach_track} in vertical dimension.
Hence, defender wins as soon as it wins the horizontal game.

\textit{{\textbf{Proposition 3:}
            If  $\mathbf{x_{z}} \in \mathcal{W}_{D, z}$ with $T_{\text{goal}}(\mathbf{x^{h}_{A}}) > T_\text{capture}(\mathbf{x^{z}})$ and $\mathbf{x}_{h} \in \mathcal{B}_{h}$, the defender is guaranteed to win.
        }}

\textit{Proof:}
Because $\mathbf{x}_{h} \in \mathcal{B}_{h}$, at all times forward it is true that $||\mathbf{p}^{h}_A - \mathbf{p}^{h}_D||_{2} \leq d_{h}$ if the defender applies optimal tracking control in Eq.
\eqref{eq:opt_d_track_horizontal} and avoids obstacles at all times.
Furthermore, because $T_{\text{goal}}(\mathbf{x^{h}_{A}}) > T_\text{capture}(\mathbf{x^{z}})$,  at $t =  T_\text{capture}(\mathbf{x^{z}})$, the attacker must have not arrived at the target goal region horizontally while it is true that $|\mathbf{p}^{z}_A - \mathbf{p}^{z}_D| \leq d_z$ if defender applies optimal reaching control in Eq.
\eqref{eq:opt_reach_vertical_ctrl}, which now means that the defender wins the game.

Note that the results of both above propositions are possible because of the construction of invariant capture set for reachability formulation introduced in Section \textbf{V}.
If we replace the winning regions $\mathcal{B}_z$, and $\mathcal{B}_h$ with the original winning region and capture set in section \textbf{III}.C, these propositions will no longer hold true.
Next, we will then combine the results of both propositions to prove the following theorem.

\textit{\textbf{Theorem:}}
    If $\mathbf{x_{z}} \in \mathcal{W}_{D, z}$ with $T_{\text{goal}}(\mathbf{x^{h}_{A}}) > T_\text{capture}(\mathbf{x^{z}})$,  and $\mathbf{x_{h}} \in \mathcal{W}_{D, h}$

\textbf{\textit{Proof:}} Since both $\mathbf{x_{z}} \in \mathcal{W}_{D, z}$ and $\mathbf{x_{h}} \in \mathcal{W}_{D, h}$.
We will consider the following 2 cases: 

\textit{Case 1: } If  $T_\text{capture}(\mathbf{x^{z}}) \leq  T_\text{capture}(\mathbf{x^{h}})$, at time $t = T_\text{capture}(\mathbf{x^{z}})$,  it must be true that $\mathbf{x}^{rel}_{z} \in \mathcal{B}_{z}$ and $\mathbf{x_{h}} \in \mathcal{W}_{D, h}$ if the defender applies optimal reaching control in both direction using Eq.
\eqref{eq:opt_reach_horizontal_ctrl}, \eqref{eq:opt_reach_vertical_ctrl}.
By \textbf{Proposition 2}, the defender is guaranteed to win.

\textit{Case 2: }
If $T_\text{capture}(\mathbf{x^{z}}) > T_\text{capture}(\mathbf{x^{h}})$,  at time $t = T_\text{capture}(\mathbf{x^{h}})$ it must be true that $\mathbf{x}_{h} \in \mathcal{B}_{h}$ and $\mathbf{x_{z}} \in \mathcal{W}_{D, z}$ if the defender applies optimal reaching control in both direction using Eq.
\eqref{eq:opt_reach_horizontal_ctrl}, \eqref{eq:opt_reach_vertical_ctrl}.
By \textbf{Proposition 3}, the defender is guaranteed to win.

In this section, we have proven the initial conditions for the defender to be guaranteed to win the original 9D game when applying decomposed reach-track-avoid control strategies in
Algorithms \ref{alg:vertical_reach_track} and \ref{alg:horizontal_reach_avoid_track}. These initial conditions forming the winning regions underapproximate the true winning regions of the original 9D game,
and hence can be smaller or more conservative than the true optimal solution, which is intractable to compute.
\section{Experiments}
\subsection{Numerical Simulations}
We perform numerical simulation to verify the properties of our proposed control algorithms and visualizing the winning regions for each player.
The parameters of the dynamic in Eq.
\eqref{eq:3D_modeling} are as follows: $k_z = 1.5, k_x = k_y = 0.7$ with the  velocity bounds $\mathbf{U}^{h}_{D} = 6$m/s and $\mathbf{U}^{h}_{A} = 3$m/s, $\mathbf{U}^{z}_{D} = 4$m/s, and $\mathbf{U}^{z}_{A} = 2 $m/s.
With these parameters, the defender's maximum speed is twice as fast as the attacker in both the horizontal and vertical directions.
All the value functions $V_{z, \infty}, V_{h, \infty}, \Phi_{z, \infty}, \Phi_{h, \infty}$ functions are then computed numerically using the OptimizedDP toolbox \cite{optimized}.

\subsubsection{Vertical Invariant Capture Set}
\begin{figure}
    \centering
    \includegraphics[scale=0.6]{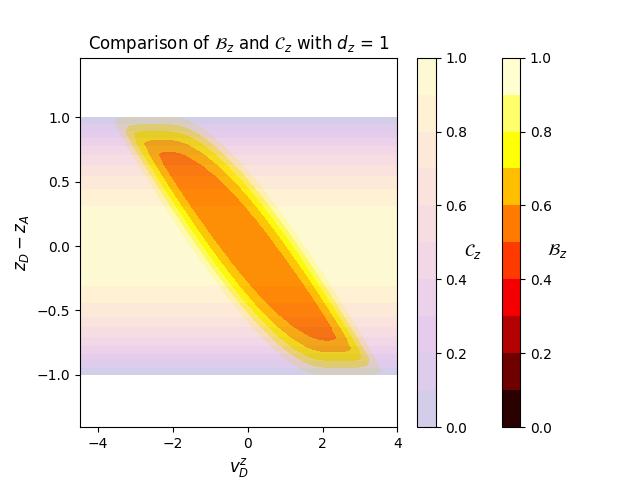}
    \caption{Plot of $z_{\text{rel}}$ against $v^{z}_{D}$, showing contour of relative distance function (light purple) and $V_{z, \infty}$ (bright yellow-orange)}
    \label{fig:B_z and C_z plot}
\end{figure}
First we compute and analyze the invariant capture set in the vertical direction $\mathcal{B}_z$ discussed in section \textbf{IV}.A.
The value function $V_{z}$ is computed over the domain of $z_{\text{rel}} \in [-10, 10] \text{ meters}, v^{z}_{D} \in [ -4., 4.
]$ m/s on a $240 \times 100$ grid.
With the set of parameters described above, the defender is powerful enough to drive the value function $V_{z}$ to 
converge as $t \rightarrow \infty$.
In the vertical game, we choose the capture distance $d_z = 1$ for which there exists a nonempty invariant set $\mathcal{B}_{z}$.
In Fig. \ref{fig:B_z and C_z plot}, we plot the contour of the relative distance function $\abs{z_D -z_A}$ and the maximum distance
value function $V_{z, \infty}$.
In this case, the collection of all level sets that are within $d_z$ from each function form the capture set $\mathcal{C}_z$ and
the invariant set $\mathcal{B}_z$.
As expected, the invariant capture set $\mathcal{B}_{z}$ for the defender is a subset of the original capture set $\mathcal{C}_{z}$.

\begin{figure}[H]
    \centering
    \begin{subfigure}{.45\textwidth}
        \centering
        \includegraphics[scale=0.5]{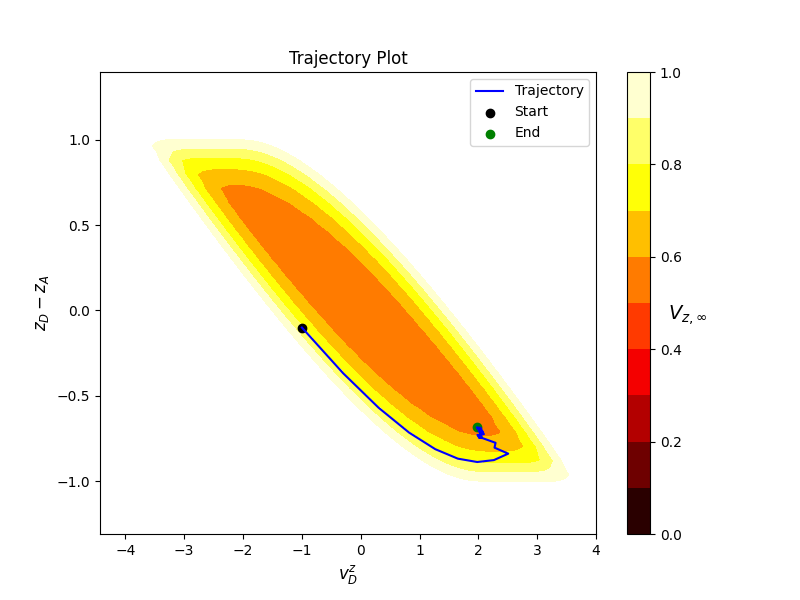}
        \caption{}
        \label{fig:vertical_tracking_subplot1}
    \end{subfigure}%
    \begin{subfigure}{.45\textwidth}
        \centering
        \includegraphics[scale=0.5]{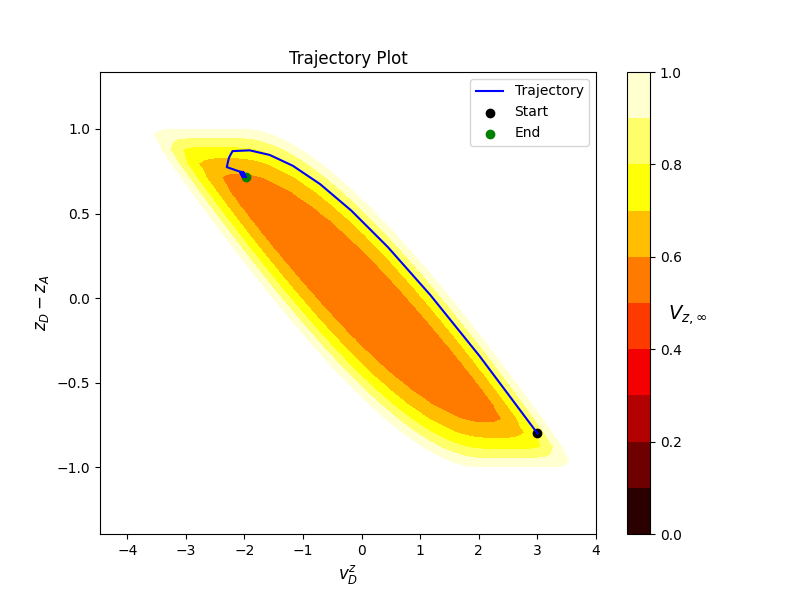}
        \caption{}
        \label{fig:vertical_tracking_subplot2}
    \end{subfigure}
    \caption{The joint state of attacker and defender remain inside the invariant capture set if the defender applies optimal tracking control}
    \label{fig:vertical_tracking_plots}
\end{figure}

Next, we simulate the behavior of the attacker and defender inside the invariant capture set $\mathcal{B}_{z}$ where both attacker and defender apply optimal control using Eq.~\eqref{eq:opt_tracking_vertical}.
Throughout the numerical simulation, the joint trajectories of both players remain inside the set and the defender can maintain relative distance $d_z$ to the attacker over time, which are shown in Fig. \ref{fig:vertical_tracking_plots}.

\subsubsection{Vertical Reach-Track Game}

\begin{figure}[H]
    \centering
    \includegraphics[scale=0.42]{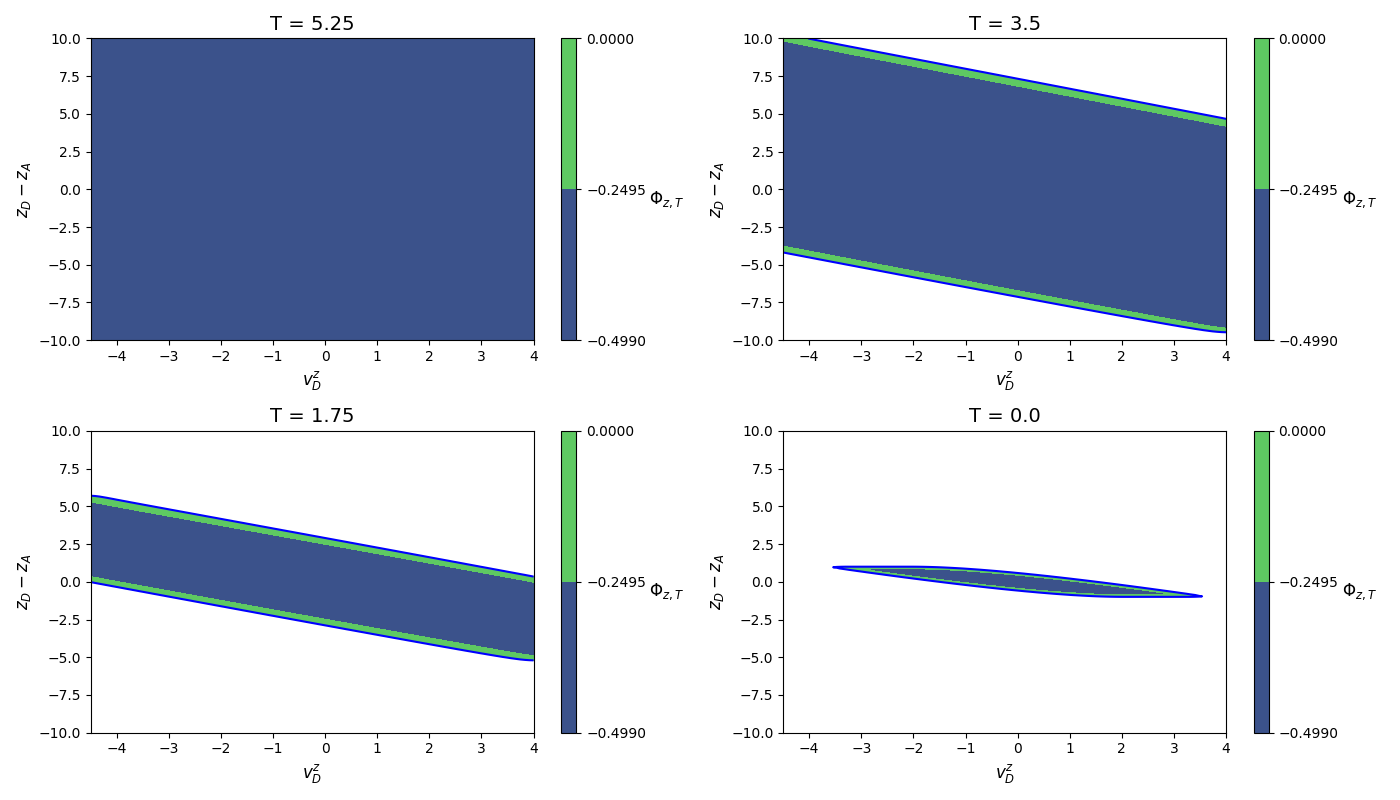}
    \caption{Winning region $\mathcal{W}_{D, z}$ for the defender (blue + green) at different durations $T$}
    \label{fig:vertical_BRT_at_different_time}
\end{figure}

\begin{figure}
    \begin{subfigure}{.5\textwidth}
        \centering
        \includegraphics[scale=0.46]{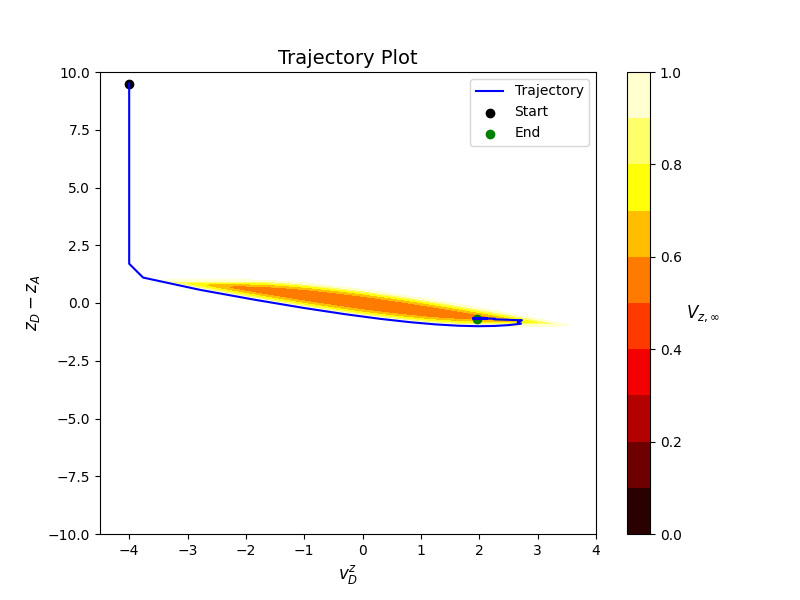}
        \caption{}
        \label{fig:vertical_BRT1}
    \end{subfigure}
    \begin{subfigure}{.5\textwidth}
        \centering
        \includegraphics[scale=0.46]{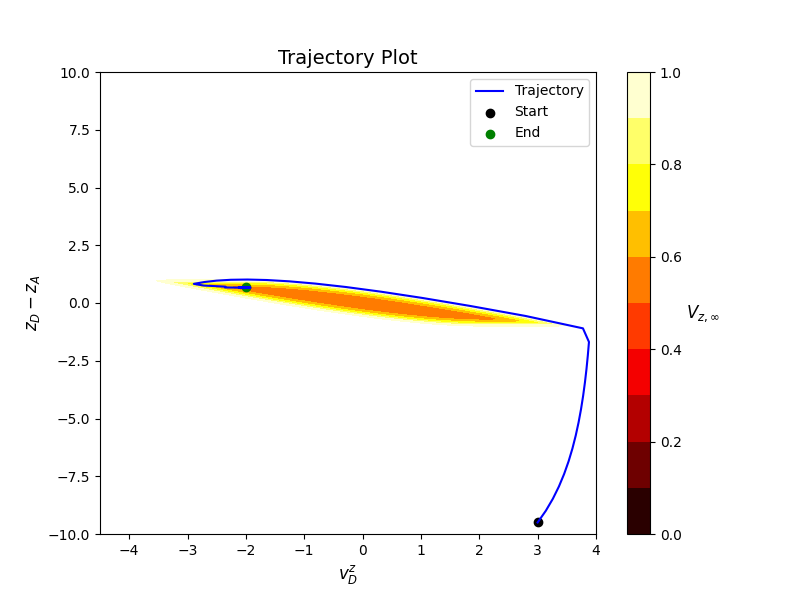}
        \caption{}
        \label{fig:vertical_BRT2}
    \end{subfigure}
    \caption{Using reach-track control algorithm, the joint system can be driven into invariant set $\mathcal{B}_{z}$. }
    \label{fig:vertical_BRT}
\end{figure}

We then use the computed $\mathcal{B}_{z}$ as the target set for computing the winning region $\mathcal{W}_{D, z}$.
Since the defender's control input $v_D^{z}$ is twice as fast as the attacker control input, given sufficent time, the defender can capture the attacker given any initial relative vertical position, 
which is shown in Fig. \ref{fig:vertical_BRT_at_different_time}.
Next, we numerically simulate the joint trajectories for the vertical sub-games at different intial relative positions, where the defender's controller follows the Algorithm \ref{alg:vertical_reach_track} and the attacker applies optimal adversarial control.
In the first scenario (Fig. \ref{fig:vertical_BRT}a), initially, the defender is above the attacker with maximum downward velocity, which it maintains to minimize the time to get close to the capture set, then it slowly starts increasing velocity to arrive at the invariant $\mathcal{B}_z$.
In Fig. \ref{fig:vertical_BRT}b, the defender is initially below the attacker and with an upward velocity of 3 m/s; it then accelerates to maximum velocity to minimize time to arrive $\mathcal{B}_{z}$ and decelerates when getting close to $\mathcal{B}_{z}$.
From both cases, once inside $\mathcal{B}_{z}$, optimal tracking controller will keep the joint state inside the capture set over time.
The numerical simulation uses forward Euler to integrate the joint state system over time given the control input.


\subsubsection{Horizontal Invariant Capture Set}
\begin{figure}
    \centering
    \includegraphics[scale=0.35]{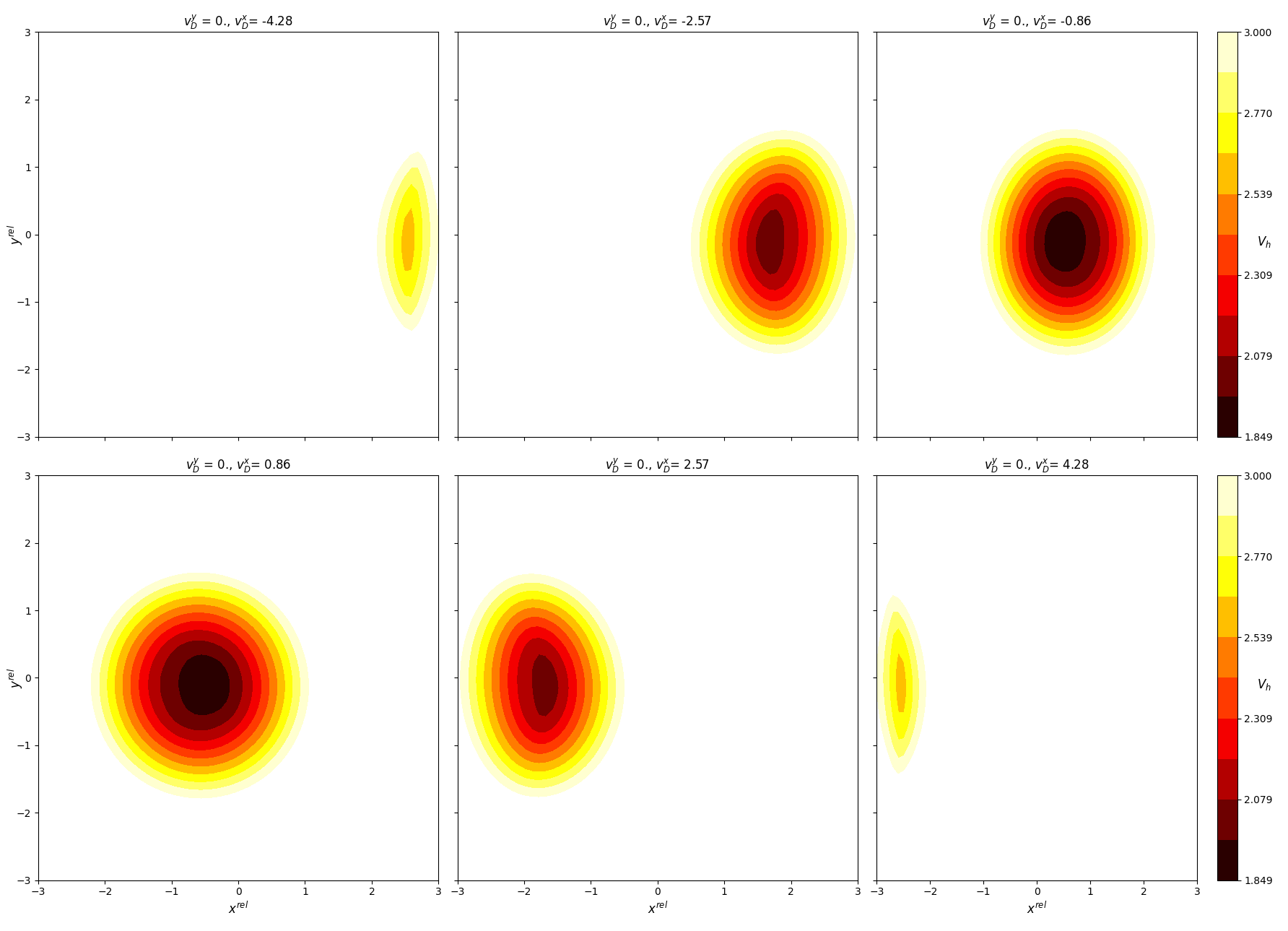}
    \caption{Contour plot of horizontal distance function $V_{h, T}$ computed at $T = 2.5$ seconds at different $v^{y}_D, v^{x}_D$}
    \label{fig:horizontal_capure_set}
\end{figure}

\begin{figure}
    \centering
    \includegraphics[scale=0.95]{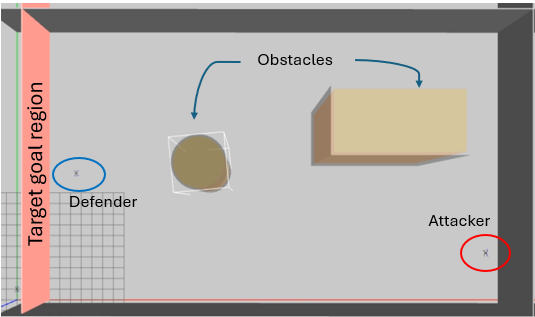}
    \caption{Top-view of our experiment configuration in Gazebo simulation}
    \label{fig:gazebo_problem_configuration}
\end{figure}
In this section, we compute the maximum distance function in the horizontal case $V_{h}$ in Eq. \eqref{eq:V_h_rel_only} over the domain of $x^{\text{rel}} \in [-3, 3]$, 
$y^{\text{rel}} \in [-3, 3] \text{ meters}$, $v^{x}_{D}, v^{y}_{D} \in [-6., 6.] \text{ m/s}$ for a grid of size 60 x 60 x 75 x 75.
With the set of parameters presented at the beginning of the section, however, $V_{h, T}$ does not converge as $T \rightarrow \infty$.
Instead, we compute $V_{h}$ for $T = 2.5$ seconds and extract the set $\mathcal{B}_{h}$ with capture radius $d_{h} = 3.0 $ ,
which is plotted at different velocity indices in Fig. \ref{fig:horizontal_capure_set}.
In this case, the defender is guaranteed to maintain a distance of $d_{h}$ and capture the attacker over a minimum time horizon of $T= 2.5$ seconds, but not indefinitely, as the defender's acceleration is not powerful enough.
As shown in \cite{FastTrackV2}, if the attacker acts suboptimally, using a time-varying $V_{h, T}$ can lead to a smaller relative distance over a longer time horizon if the defender applies optimally.
In practice, this can be true as the attacker might not be as physically powerful as a single integrator that is modeled here.

\begin{figure}
    \begin{subfigure}{.5\textwidth}
        \centering
        \includegraphics[scale=0.5]{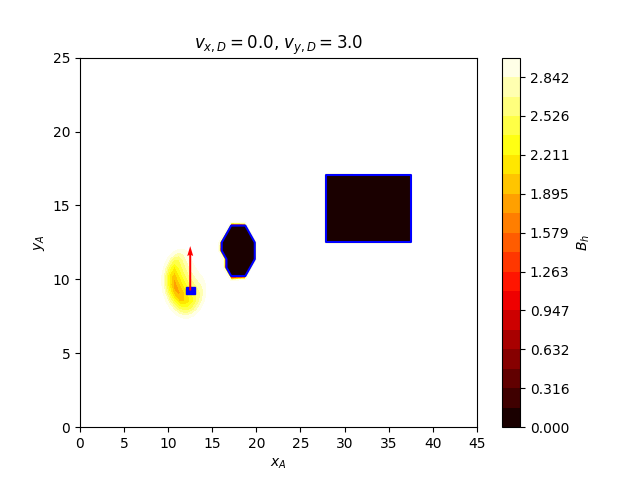}
        \caption{}
    \end{subfigure}
    \begin{subfigure}{.5\textwidth}
        \centering
        \includegraphics[scale=0.5]{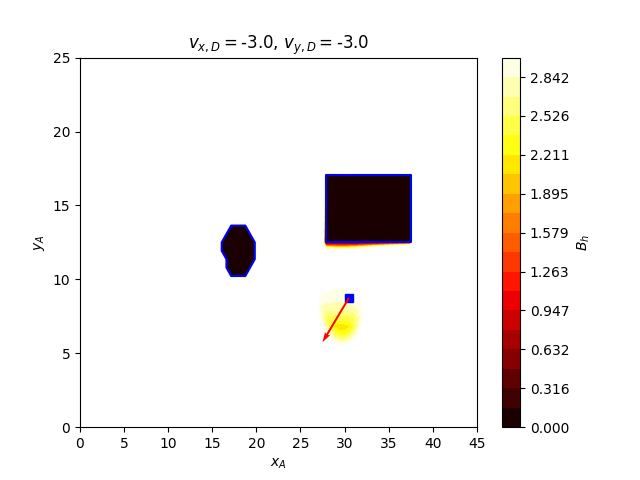}
        \caption{}
    \end{subfigure}
    \begin{subfigure}{.5\textwidth}
        \centering
        \includegraphics[scale=0.5]{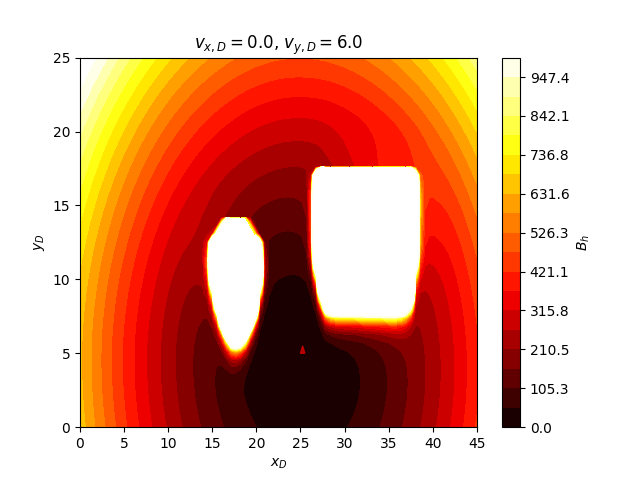}
        \caption{}
    \end{subfigure}
    \begin{subfigure}{.5\textwidth}
        \centering
        \includegraphics[scale=0.5]{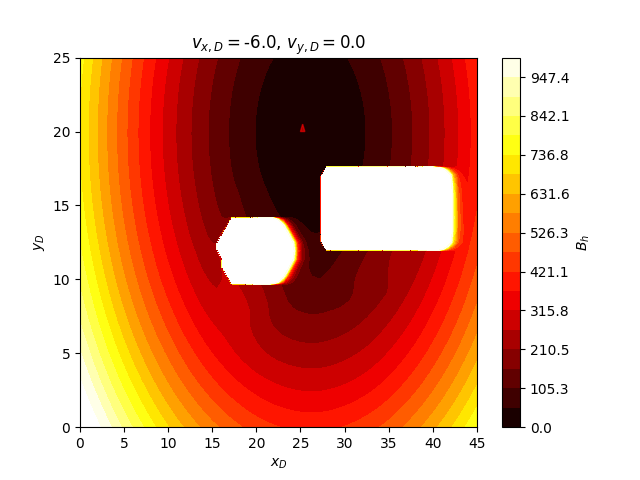}
        \caption{}
    \end{subfigure}
    \caption{ Contour of the value function $V_h$ }
    \label{fig:horizontal_invariant_set_with_obs}
\end{figure}

\begin{figure}
    \begin{subfigure}{1\textwidth}
        \centering
        \includegraphics[scale=0.45]{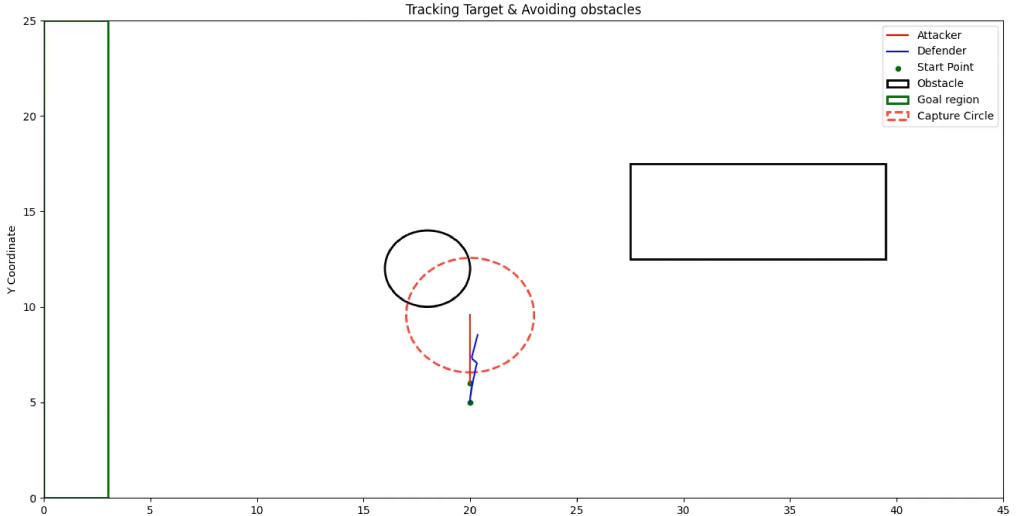}
        \caption{$T=1$}
    \end{subfigure}
    \begin{subfigure}{1\textwidth}
        \centering
        \includegraphics[scale=0.45]{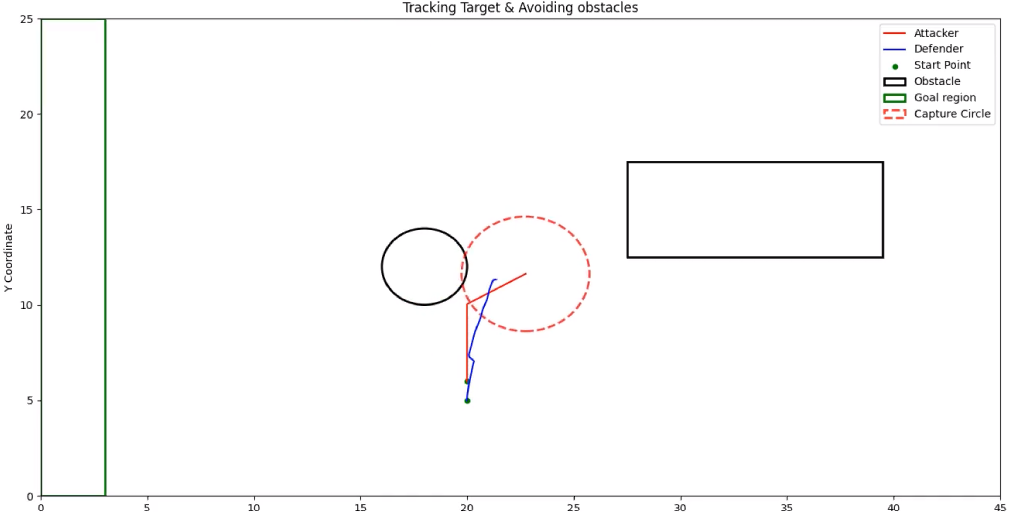}
        \caption{$T=2$}
    \end{subfigure}
    \begin{subfigure}{1\textwidth}
        \centering
        \includegraphics[scale=0.45]{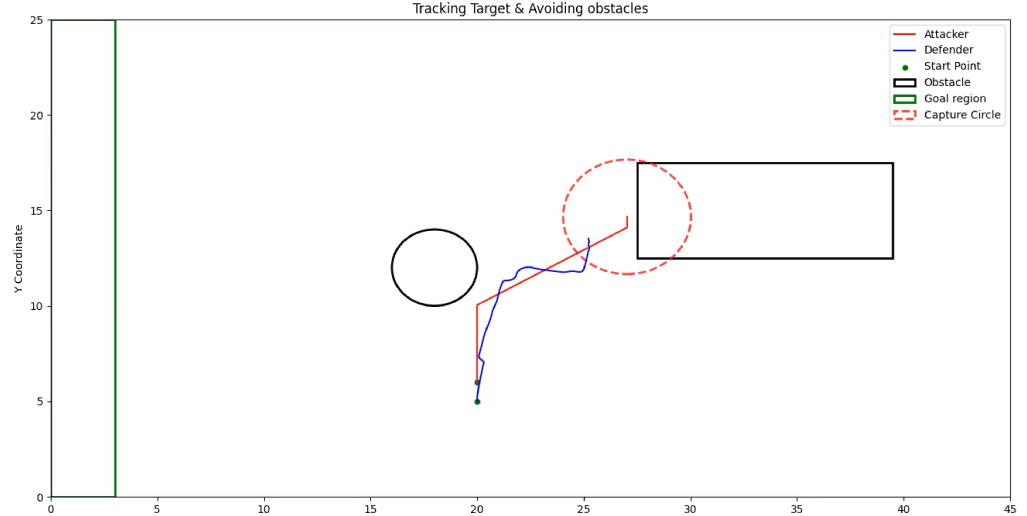}
        \caption{$T=3$}
    \end{subfigure}
    \begin{subfigure}{1\textwidth}
        \centering
        \includegraphics[scale=0.45]{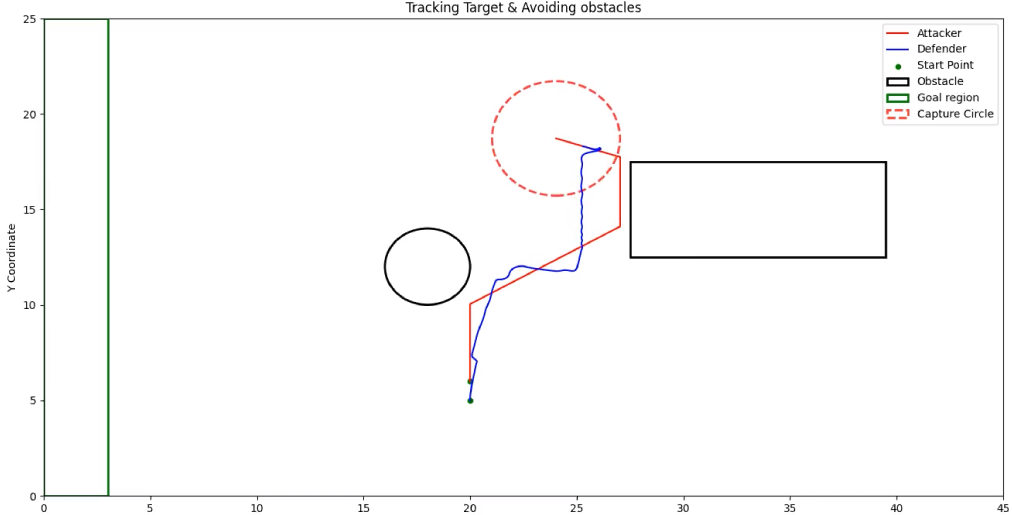}
        \caption{$T=4$}
    \end{subfigure}
    \caption{Simulation of the defender applying optimal tracking control to track attacker and avoid obstacles}
    \label{fig:track_avoid_simulate_plot}
\end{figure}

Next, we will compute and simulate the horizontal value function in Eq. \eqref{eq:horizontal_value_func_obs}, which is defined over the domain $(x_{A}, x_{D}) \in [0, 45] $ meters, $(y_A, y_D) \in [0, 25]$ meters and $(v^{x}_D, v^{y}_D) \in [-6., 6.]$ meter/second on a grid of $(x_{A}, y_{A}, x_D, y_D, v^{x}_{D}, v^{y}_{D} ) $ with size 60 x 30 x 75 x 36 x 13 x 13.
The obstacles configuration are shown in our Gazebo simulation configuration in Fig. (\ref{fig:gazebo_problem_configuration}).
The tracking is now more challenging for the defender, as it is modeled as a double integrator. If we compute $\mathcal{B}_h$, the capture set will be empty as the defender is not agile enough to maintain tracking distance of $d_h = 3.$ while avoiding obstacles. Hence, for the sake of experiments, we adjust the parameter $k_x=k_y=2$ and reduce maximum horizontal speed of attacker $\mathbf{U}^{h}_{A} = 2$m/s. Different slices of the value function are shown in Fig. \ref{fig:horizontal_invariant_set_with_obs}. The top row of the figure shows the invariant capture set given the current defender position and velocity. The last row of the figure show contour of the value function given the fixed attacker position. It can be observed in the last row the inflated region around the original obstacles corresponds to the BRT of the defender, which matches with our tracking distance formulation.
We also simulate the track-avoid game where the defender applies optimal control to track the attacker within $d_h=3.0$ m while avoiding obstacle at the same time. The simulation is shown in Fig. \ref{fig:track_avoid_simulate_plot}.

\subsubsection{Horizontal Reach-Avoid Game}
Similar to the vertical case, we then use $\mathcal{B}_{h}$ as the target set to compute $\Phi_{h}$ and obtain the winning region $\mathcal{W}_{D, h}$,  $\mathcal{W}_{A, h}$ for the horizontal game, which is defined over the domain $(x_{A}, x_{D}) \in [0, 45] $ meters, $(y_A, y_D) \in [0, 25]$ meters and $(v^{x}_D, v^{y}_D) \in [-6., 6.]$ meter/second on a grid of $(x_{A}, y_{A}, x_D, y_D, v^{x}_{D}, v^{y}_{D} ) $ with size 85 x 45 x 85 x 45 x 8 x 7.
The target set and obstacles configuration are shown in our Gazebo simulation configuration in Fig. \ref{fig:gazebo_problem_configuration}.
In our configuration, the target region $\mathcal{T}$ is a region behind an invisible vertical wall (red wall in Fig.~\ref{fig:gazebo_problem_configuration}) 
defined as $\mathcal{T} = \{ \mathbf{p} \in \mathbb{R}^3 \mid x \leq 3$ \}.
Solving for the game value $\Phi_{h}$ takes in total 3 hours for $T = 22$ seconds using the OptimizedDP toolbox \cite{optimized}.

\begin{figure}
    \begin{subfigure}{.5\textwidth}
        \centering
        \includegraphics[scale=0.5]{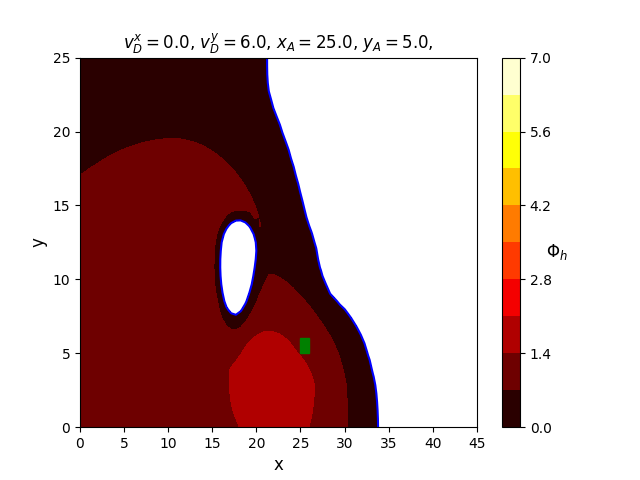}
        \caption{}
        \label{fig:horizontal_BRT1}
    \end{subfigure}
    \begin{subfigure}{.5\textwidth}
        \centering
        \includegraphics[scale=0.5]{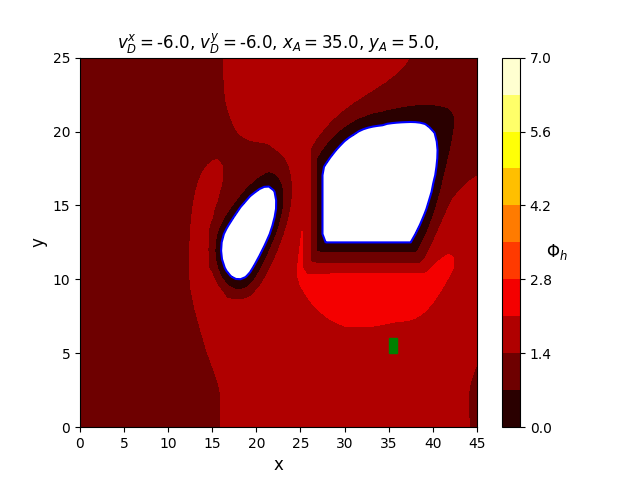}
        \caption{}
        \label{fig:horizontal_BRT2}
    \end{subfigure}
    \begin{subfigure}{.5\textwidth}
        \centering
        \includegraphics[scale=0.5]{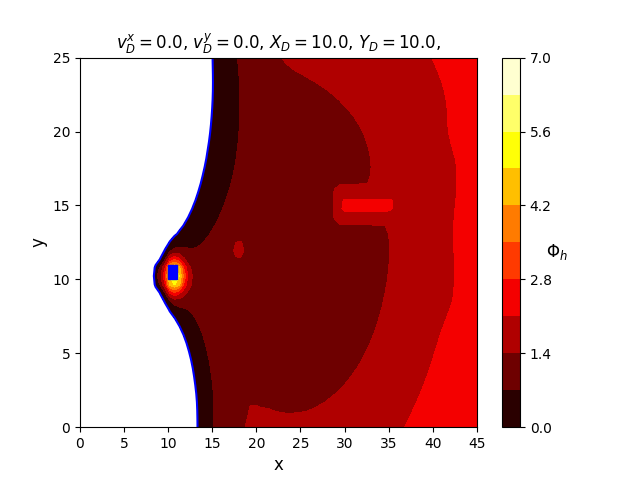}
        \caption{}
        \label{fig:horizontal_BRT3}
    \end{subfigure}
    \begin{subfigure}{.5\textwidth}
        \centering
        \includegraphics[scale=0.5]{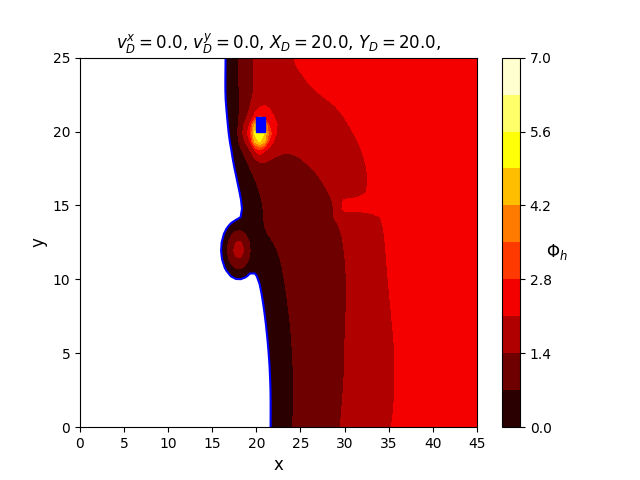}
        \caption{}
        \label{fig:horizontal_BRT4}
    \end{subfigure}
    \caption{Winning region of the defender (red) and of the attacker (white) given fixed defender (blue) and attacker (green)'s states. }

    \label{fig:horizontal_BRTs}
\end{figure}

Since the function $\Phi_h$ is a 6D value function, it is not possible to visualize it and the winning regions $\mathcal{W}_{D, h}$,
$\mathcal{W}_{A, h}$ in a single plot. 
Instead, we will attempt to do so by slicing the 6D value function 
into 2D slices. In the top row of 
Fig. \ref{fig:horizontal_BRTs}, we fix the velocity of the defender and position of the attacker ( green rectangle), 
and plot the contour of the 2D value function $\Phi_{h}(x^h_{D}, y^{h}_D)$. In this row, the red region is the super-zero level set of $\Phi_{h}(x^h_{D}, y^{h}_D)$,
which contains all the positions where defender is guaranteed to win the game and the white region contains all 
the defender's positions leading it to lose the game by either hitting obstacles or not be able capture the attacker.
Note that the obstacles region in white inflates a bit more due to inertia and acceleration bound of the defender.
Similarly,
in the bottom row of Fig. \ref{fig:horizontal_BRTs}, we fix the position of the defender (blue rectangle) and its velocity,
 and plot the contour of the 2D value function $\Phi_{h}(x^h_{A}, y^{h}_A)$. In this case, the red region contains all the attacker positions that
 will result in a win for the defender and the white region contains all the attacker positions that will result in a win for the attacker.

\begin{figure}[H]
    \begin{subfigure}{.5\textwidth}
        \centering
        \includegraphics[scale=0.34]{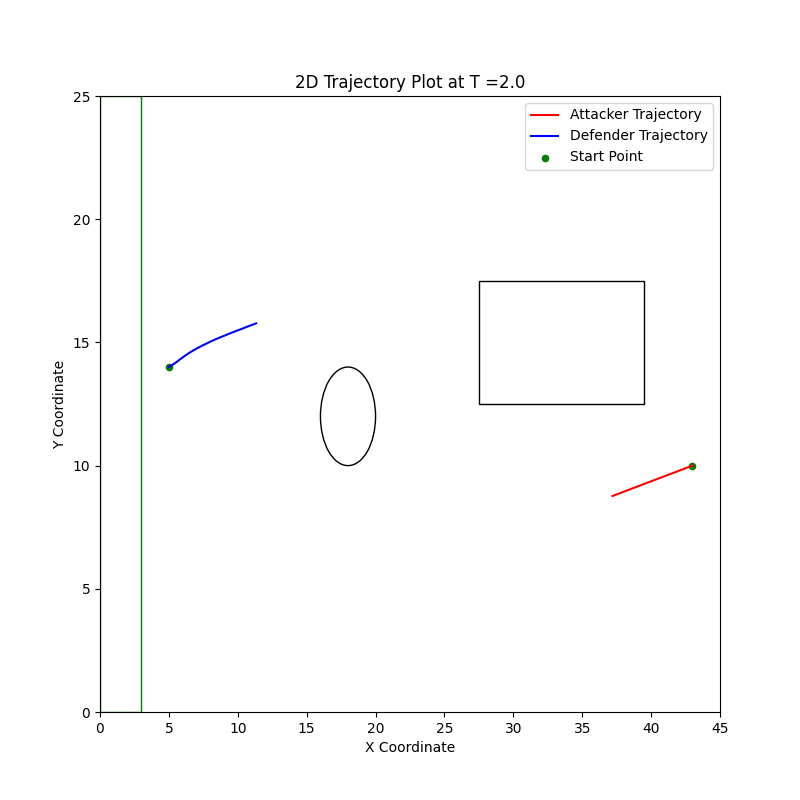}
        \caption{T = 2 seconds}
        \label{fig:horizontal_numerical_simulation1}
    \end{subfigure}
    \begin{subfigure}{.5\textwidth}
        \centering
        \includegraphics[scale=0.34]{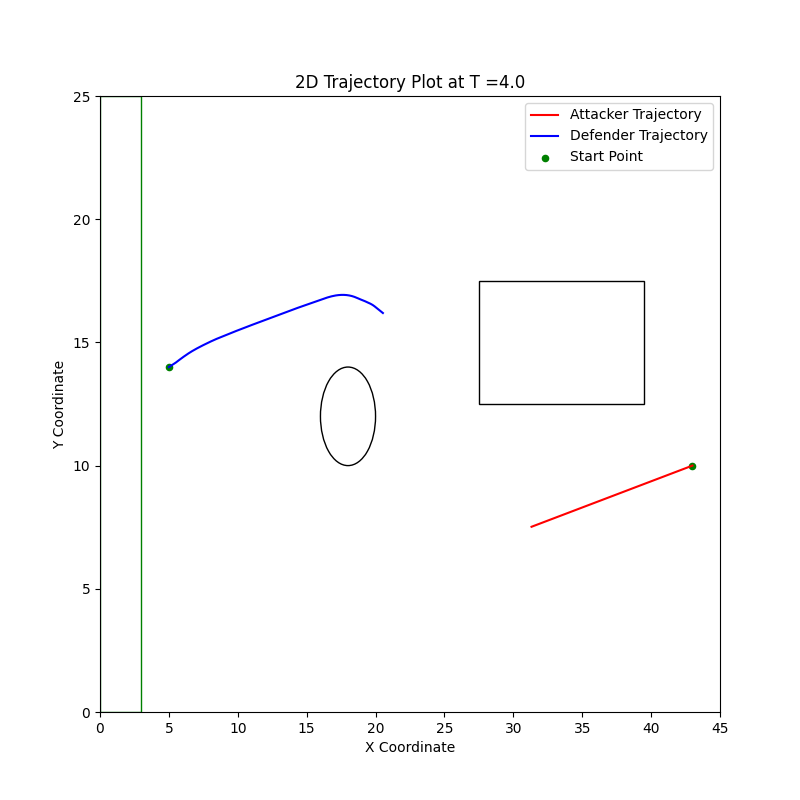}
        \caption{T = 4 seconds}
        \label{fig:horizontal_numerical_simulation2}
    \end{subfigure}
    \begin{subfigure}{.5\textwidth}
        \centering
        \includegraphics[scale=0.34]{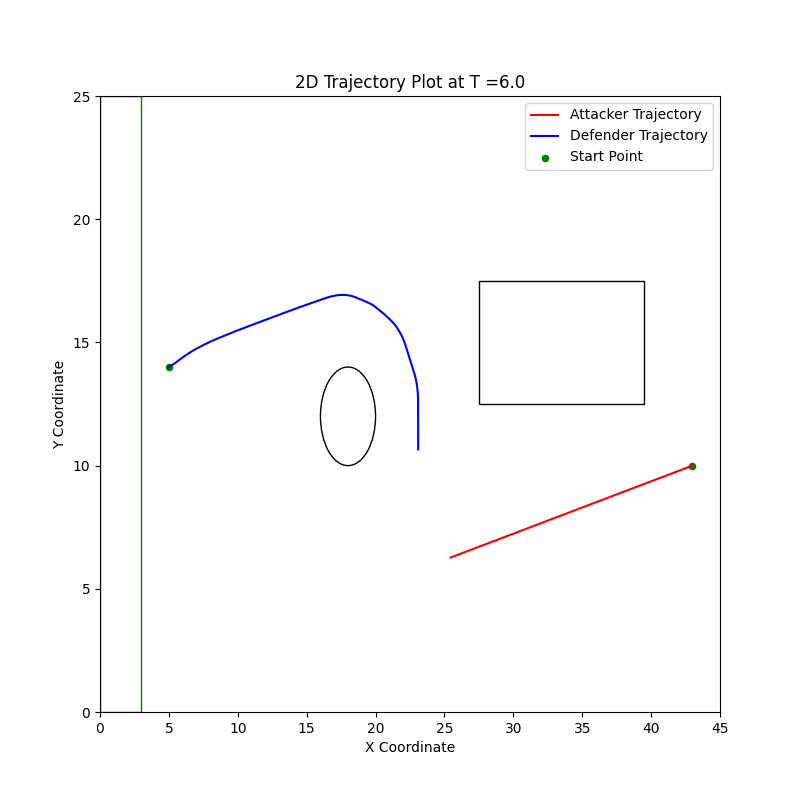}
        \caption{T = 6 seconds}
        \label{fig:horizontal_numerical_simulation3}
    \end{subfigure}
    \begin{subfigure}{.5\textwidth}
        \centering
        \includegraphics[scale=0.34]{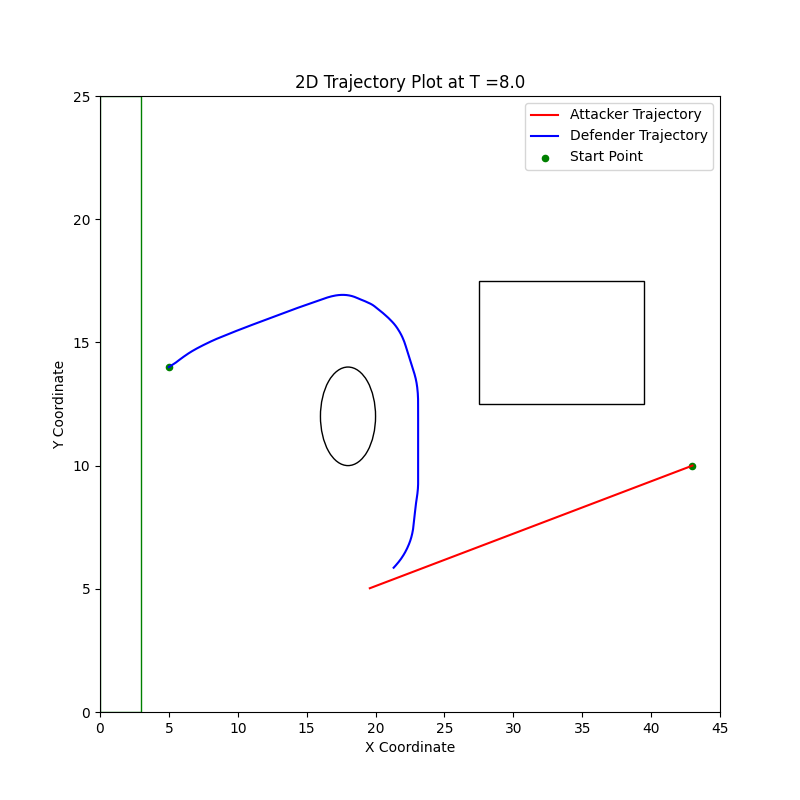}
        \caption{T = 8 seconds}
        \label{fig:horizontal_numerical_simulation4}
    \end{subfigure}
    \begin{subfigure}{.5\textwidth}
        \centering
        \includegraphics[scale=0.34]{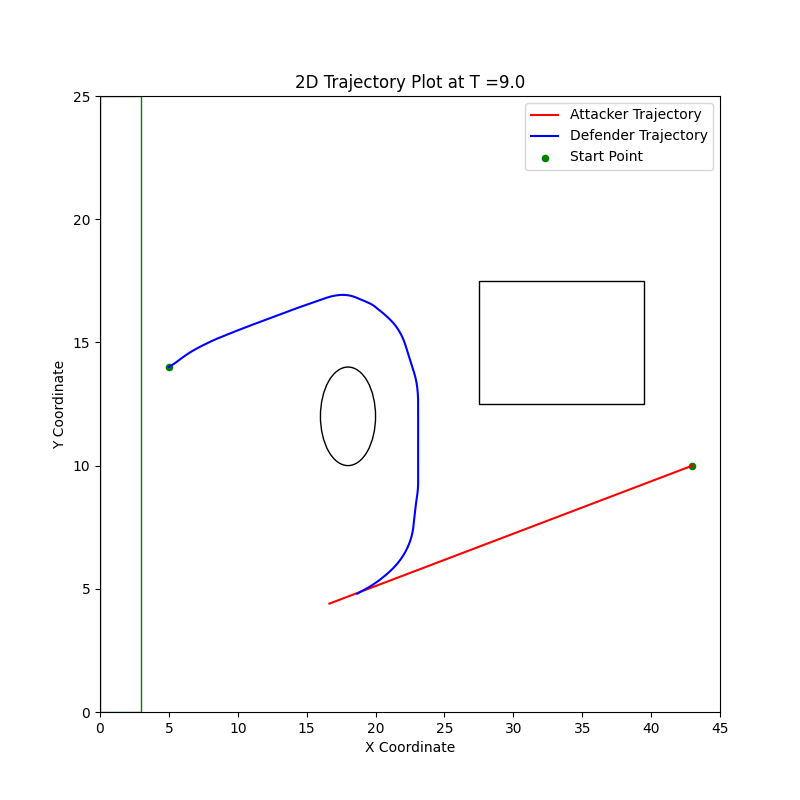}
        \caption{T = 9 seconds}
        \label{fig:horizontal_numerical_simulation5}
    \end{subfigure}
    \begin{subfigure}{.5\textwidth}
        \centering
        \includegraphics[scale=0.34]{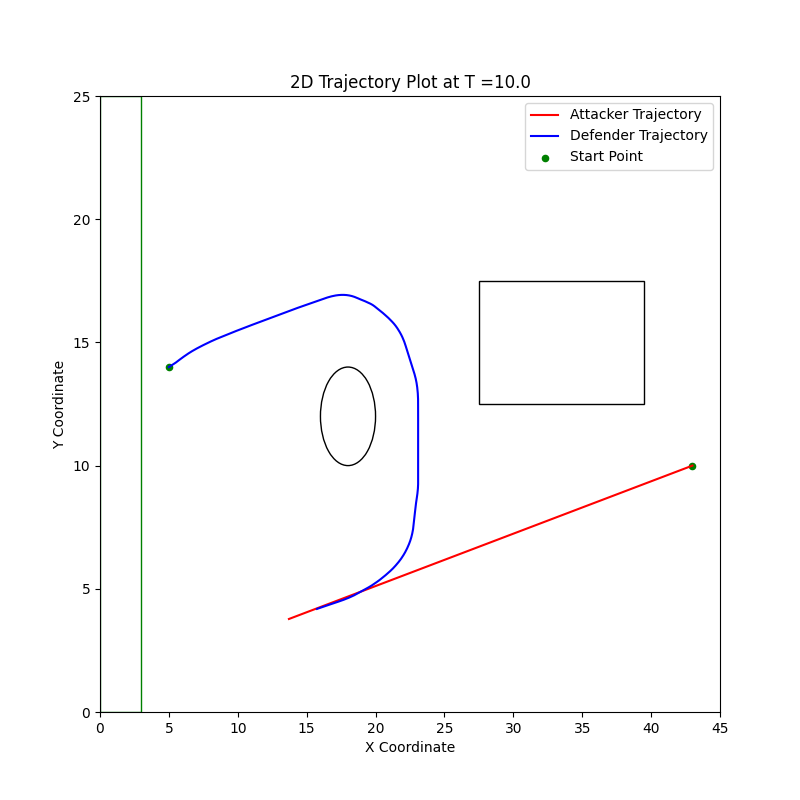}
        \caption{T = 10 seconds}
        \label{fig:horizontal_numerical_simulation6}
    \end{subfigure}
    \caption{Numerical simulation of the joint trajectories in the horizontal game at different time steps.} \label{fig:horizontal_numerical_simulation}

\end{figure}

Using the computed horizontal game value $\Phi_{h,T}$, we then simulate the joint trajectories in the horizontal reach-avoid game (shown in Fig.~\ref{fig:horizontal_numerical_simulation}.)
In Fig. \ref{fig:horizontal_numerical_simulation}, the attacker (red) is moving towards target region (left vertical green line)
while the defender (blue) is trying to capture the attacker while avoiding the obstacles (circle and rectangular blocks).
As described in Algorithm \ref{alg:horizontal_reach_avoid_track}, initially, the defender applies optimal reaching control to arrive at $\mathcal{B}_{h}$ and capture the attacker.
From there, the defender switches to using the optimal tracking controller to maintain the capture status thereafter.
  It can be observed in Fig. \ref{fig:horizontal_numerical_simulation5} that at $T=9$, the defender is able to close the distance, arrive at $\mathcal{B}_{h}$ and capture the attacker.
The defender then switch to use the horizontal tracking controller to track the attacker, and hence, at $T=10$, the defender is still able to 
maintain the relative distance $d_h$.

\subsection{Gazebo Simulation}

\begin{figure}[H]   
    \centering
    \includegraphics[scale=0.6]{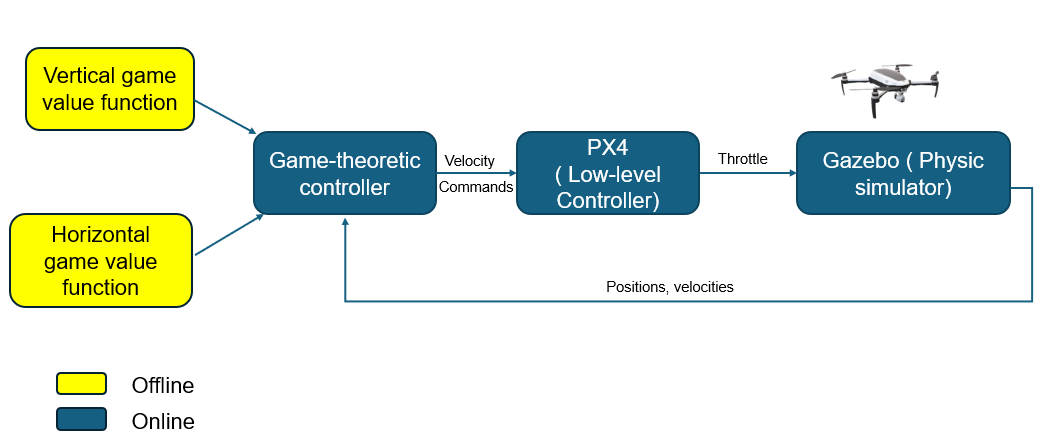}
    \caption{Software configuration of our experiment in Gazebo simulation.}
    \label{fig:software_configuration}
\end{figure}

Finally, we combine both Algorithms \ref{alg:vertical_reach_track} and \ref{alg:horizontal_reach_avoid_track} into quadrotor systems in the Gazebo physics simulator.
The software configuration for our experiment is shown in Fig. \ref{fig:software_configuration}.
In this experiment, all information about the state of each agent is assumed to be perfectly known and
obtained from Gazebo ground truth.
PX4 Ardupilot is used as the interface to translate optimal velocity commands to throttle commands that control the quadrotor.
The configuration of our reach-avoid game is shown in Fig. \ref{fig:gazebo_problem_configuration}.

\begin{figure}[H]
    \begin{subfigure}[b]{\textwidth}
        \centering
        \includegraphics[angle=90,origin=c, scale=0.4]{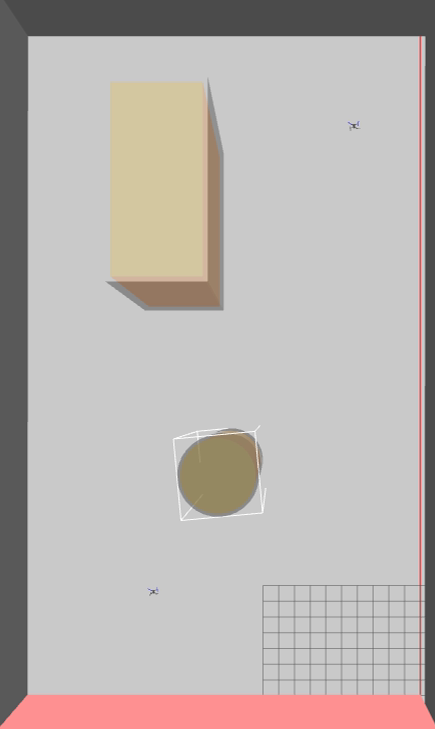}
        \hfill
        \includegraphics[scale=0.58]{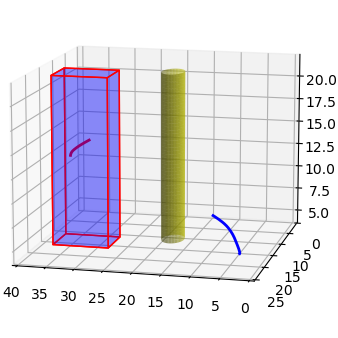}
        \caption{T = 1 second}
        \label{fig:3d_capture_track1}
    \end{subfigure}
    \begin{subfigure}[b]{\textwidth}
        \centering
        \includegraphics[angle=90,origin=c, scale=0.4]{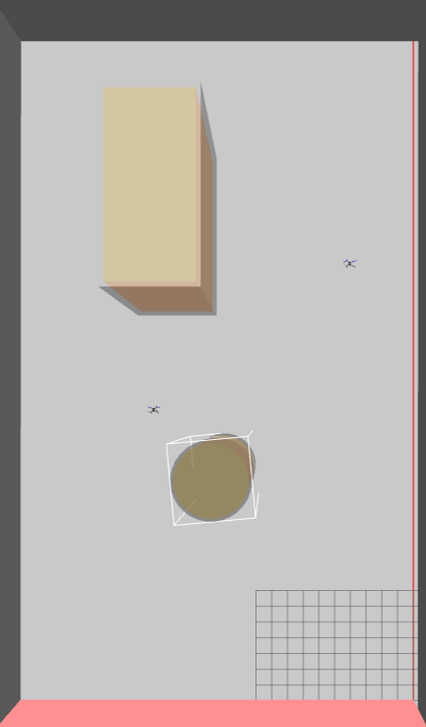}
        \hfill
        \includegraphics[scale=0.58]{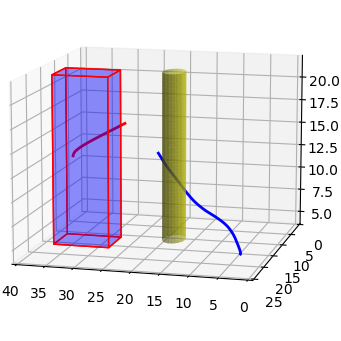}
        \caption{T = 2 seconds}
        \label{fig:3d_capture_track2}
    \end{subfigure}
    \begin{subfigure}{\textwidth}
        \centering
        \includegraphics[angle=90,origin=c, scale=0.4]{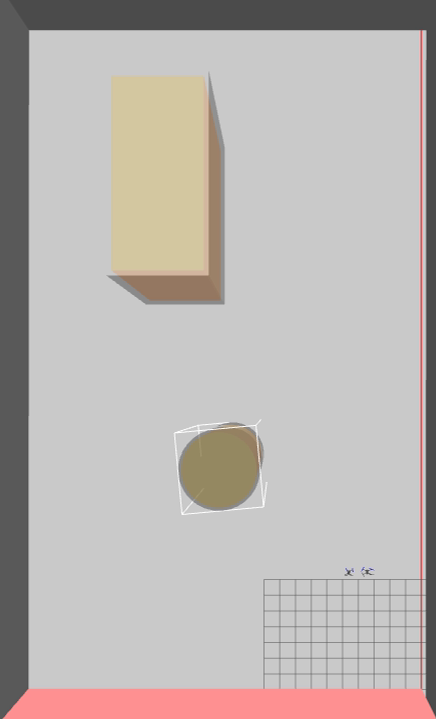}
        \hfill
        \includegraphics[scale=0.58]{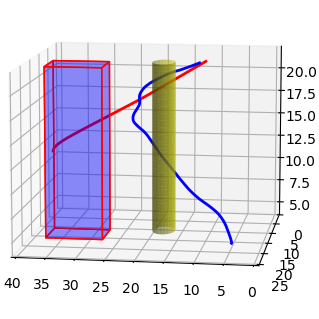}
        \caption{T = 5 seconds}
        \label{fig:3d_capture_track3}
    \end{subfigure}
    \caption{Positions (left column) and past trajectories (right column) of defender (blue) and attacker (red) over time in Gazebo.}
    \label{fig:3d_capture_track_gazebo}
\end{figure}

As shown in Fig. \ref{fig:3d_capture_track1}, initially, the attacker is above the defender, and it constantly moves up at a constant velocity of 2 m/s.
The vertical reach-track algorithm helps the defender arrive at the same height as the attacker and maintain the capture status.
In this particular configuration, the defender captures the attacker in the vertical direction first and then wins the horizontal game later.
Eventually, the defender is able to capture the attacker in both horizontal and vertical games.
It can be seen from the simulation in Gazebo that our algorithm can effectively capture the attacking drone in three dimensional space.

\section{Conclusion}
In this paper, we have proposed a decomposition method that breaks down the 9-dimensional reach-avoid game into horizontal and vertical sub-games,
each of which is tractably solved using HJ reachability analysis. To ensure capture guarantees when combining the controllers from
both sub-games, we have introduced invariant capture set as the reach set for the defender and proposed a
reach-track control algorithm in each sub-games. We have also provided analysis of the conditions under which the defender is 
guaranteed to win the original game using our proposed algorithms. Finally, we have demonstrated the effectiveness of our method through numerical simulations 
and physical experiments using quadrotors in a Gazebo simulator.
Although our method proposed is for a single defender and single attacker, it can be extended to multiple defenders 
and multiple attackers scenarios using optimization methods such as mixed integer programming
or bipartite graph matching. The winning condition, in this case, would require the
defender to win both sub-games.
Despite being effective for capturing the attacker in three-dimensional space with a double integrator model of the defender, our method is based on numerical computation, which can limit the range of spatial space and the velocity the quadrotors can operate with.
Further research into analytical methods to be combined with our method can help expand this operation range.
In addition, we also assume that all information about the attacker (velocity, state) is available to the defender, which might not always be true in the wild due to sensor error, occlusion, detection error, etc. Such problems might be mitigated by deployment of coordinated multi-agent defender systems for both sensing and capturing the attacker, which will require further investigation.

\section{Acknowledgments}
\noindent
Funding for this work was provided by the Canadian Department of National Defence.
\bibliography{references}

\end{document}